\documentclass[aps, prl, 10pt, twocolumn, superscriptaddress, noshowpacs, preprintnumbers, longbibliography, bibnotes,hyperref,floatfix]{revtex4-1}

\usepackage[colorlinks=true,breaklinks=true]{hyperref}
\hypersetup{allcolors=[rgb]{0.0 0.0 1.0},linkcolor=[rgb]{0.75 0.05 0.05}}
\usepackage[dvipsnames]{xcolor}
\usepackage{mathtools}
\usepackage{amsmath}
\usepackage{amsfonts}
\usepackage{amssymb}
\usepackage{bbold}
\usepackage{multirow}
\usepackage{bm}
\usepackage{hyperref}
\long\def\exclude#1{}
\usepackage{orcidlink}
\usepackage{comment}

\usepackage{booktabs}
\usepackage{siunitx}
\begin{document}

\title{
Small Progenitors, Large Couplings:\\
Type Ic Supernova Constraints on Radiatively Decaying Particles
}

\author{Francisco R. Cand\'{o}n~\orcidlink{0009-0002-3199-9278}}
\affiliation{Fakult\"at f\"ur Physik, TU Dortmund, Otto-Hahn-Stra{\ss}e 4, Dortmund D-44221, Germany}
\affiliation{Centro de Astropart\'iculas y F\'isica de Altas Energ\'ias (CAPA), Universidad de Zaragoza, 50009 Zaragoza, Spain}

\author{Damiano F.\ G.\ Fiorillo \orcidlink{0000-0003-4927-9850}}
\affiliation{Gran Sasso Science Institute, Viale F. Crispi 7, L’Aquila, 67100, Italy}
\affiliation{Istituto Nazionale di Fisica Nucleare (INFN), Sezione di Napoli,
Complesso Universitario di Monte Sant’Angelo, Via Cintia, 80126 Napoli, Italy}
\affiliation{Deutsches Elektronen-Synchrotron DESY,
Platanenallee 6, 15738 Zeuthen, Germany}

\author{\hbox{Hans-Thomas Janka \orcidlink{0000-0002-0831-3330}}}
\affiliation{Max-Planck-Institut f\"ur Astrophysik, Karl-Schwarzschild-Stra{\ss}e 1, 85748 Garching, Germany}

\author{\hbox{Bart F. A. van Baal  \orcidlink{0009-0001-3767-942X}}}
\affiliation{The Oskar Klein Centre, Department of Astronomy, Stockholm University, AlbaNova, SE-10691 Stockholm, Sweden}

\author{Edoardo Vitagliano
\orcidlink{0000-0001-7847-1281}}
\affiliation{Dipartimento di Fisica e Astronomia, Università degli Studi di Padova,
Via Marzolo 8, 35131 Padova, Italy}
\affiliation{Istituto Nazionale di Fisica Nucleare (INFN), Sezione di Padova,
Via Marzolo 8, 35131 Padova, Italy}

\begin{abstract}
Supernova (SN) 1987A provides classic bounds on gamma-ray flashes from the radiative decay of sub-GeV particles, but the latter may decay so rapidly as to be shielded by the stellar envelope. Using axionlike particles with photon coupling as a benchmark, we show that Type Ic core-collapse supernovae largely evade this attenuation due to their compact progenitors. We identify two regimes. At small couplings, while individual flashes may be missed by Fermi-LAT, the high Type Ic rate enables a stacking analysis of nondetections. At larger couplings, decay photons trigger fireball formation; the absence of a signal from the rare broad-lined Type Ic SN~1998bw excludes this scenario. The resulting bounds significantly exceed those from SN~1987A for short decay lengths.

\end{abstract}

\date{\today}

\maketitle

{\bf\textit{Introduction}}---Feebly interacting particles with mass smaller than the GeV scale are the target of many astrophysical, cosmological, and laboratory searches~\cite{Essig:2013lka,Agrawal:2021dbo,Abdullahi:2022jlv,Berryman:2022hds,Baryakhtar:2022hbu}. On Earth, the best way to detect these particles is through beam-dump experiments, in which a beam is shot on a target, and upon the interaction of the former with the latter novel particles might be produced~\cite{CHARM:1985anb,Riordan:1987aw,Blumlein:1990ay,NA64:2020qwq,Dolan:2017osp,Bauer:2017ris,Bauer:2018onh,Magill:2018jla,Krnjaic:2019rsv,Bolton:2019pcu,Calibbi:2020jvd,Capozzi:2023ffu,Knapen:2024fvh,Li:2025yzb}. Because of their mass and coupling, novel particles can decay radiatively back to standard model degrees of freedom. The bounds from beam-dump experiments share a universal shape featuring a floor and a ceiling, as the production of particles that interact too feebly is suppressed, but particles that interact too strongly will decay in the shield before reaching the decay volume.

\begin{figure}[t!]
    \centering
    \includegraphics[width=1.\columnwidth]{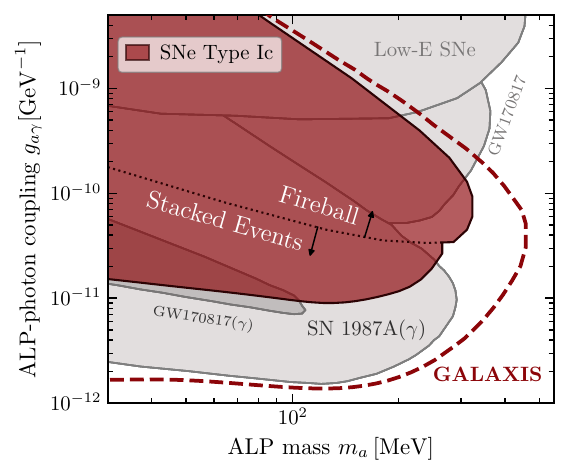}
    \caption{Novel bounds on ALP-photon coupling from Type Ic SNe from stacked-events strategy and from the nonobservation of a fireball from SN~1998bw. Existing constraints include low-energy SNe~\cite{Caputo:2022mah,Fiorillo:2025yzf}, neutron star mergers~\cite{Diamond:2023cto,Dev:2023hax}, and gamma-ray and x-ray signals at the Solar Maximum Mission and the Pioneer Venus Observatory~\cite{Jaeckel:2017tud,Caputo:2021rux,Hoof:2022xbe,Diamond:2023scc}. The red dashed line shows the projected sensitivity for a Type Ic SN at 10 Mpc with a next-generation $\gamma$-ray telescope network such as GALAXIS~\cite{Manzari:2024jns}.
    } 
    \label{fig:results_exclusion}
\end{figure}

Supernovae (SNe) are gargantuan beam-dump experiments~\cite{Raffelt:1996wa}. The hot, dense core of the protoneutron star (PNS, which has a radius $R_{\rm NS}\simeq 10\,\rm km$) serves as both beam and target for the production of radiatively decaying particles such as axions~\cite{Giannotti:2010ty,Jaeckel:2017tud,Caputo:2021rux,Caputo:2021kcv,Hoof:2022xbe,Lella:2022uwi,Lella:2024dmx,Benabou:2024jlj,Alda:2024cxn,Fiorillo:2025sln}, dark photons~\cite{Kazanas:2014mca,DeRocco:2019njg,Calore:2021lih}, and sterile neutrinos~\cite{Carenza:2023old,Akita:2023iwq}, which are emitted with energies $E \simeq 100 \,\rm MeV$. We will focus on the case of axionlike particles (ALPs) with a coupling to photons, but the conclusions we will reach  apply to any feebly interacting particle featuring decay channels with photons in the final state. Depending on the decay length $\lambda_a$ of the ALPs, they might decay outside of the stellar layers surrounding the core, so that the photon flux resulting from the decay would show up in gamma-ray~\cite{Jaeckel:2017tud,Caputo:2021rux,Hoof:2022xbe} and x-ray telescopes~\cite{Diamond:2023scc}. However, the stellar envelope can act as a shield (see Refs.~\cite{Caputo:2022mah,Fiorillo:2025yzf} for the effects of the decay within this shield) so that these bounds have a ceiling akin to the one of beam-dump searches. Sanduleak-69 202, the blue supergiant progenitor of SN~1987A, had a best-estimate radius of $R_{\rm prog}= 31$--$37\,R_\odot$, $R_\odot\simeq 7\times 10^{10}\,\rm cm$ \cite{Menon+2019,Ono+2020,Orlando+2020,Utrobin+2021}. The main point of our Letter is that the most stringent constraints for decay lengths $\lambda_a\simeq R_{\rm prog}$ 
do not come from the closest observed supernova (SN) (i.e., from SN~1987A), but are rather obtained from the most compact core-collapse SNe, even if more distant. Therefore, we show in the following that 
Type Ic SNe, which sport radii as small as $R_{\rm prog}\simeq 0.1$--$1\,R_\odot$~\cite{Tauris:2015xra}, can reliably surpass the bounds of SN~1987A and reach couplings larger by one order of magnitude, probing new portions of the parameter space (see Fig.~\ref{fig:results_exclusion}).

{\bf\textit{ALP-induced gamma rays from Type~Ic SNe}}---The inner core of Type~Ic SNe is expected to have properties very similar to those of Type~II SNe, of which SN~1987A was a member; we exclude here the broad-line SNe~Ic, whose powerful energy output is probably connected with rapid rotation of the stellar core and new-born PNS.
The PNS is then a similarly efficient factory of putative sub-GeV particles. Here, we focus on radiatively decaying species, of which the prime and most well-studied example is the ALP coupling to photons with the Lagrangian $\mathcal{L}=-\frac{1}{4}g_{a\gamma}a F \tilde{F}$, where $a$ is the ALP field, $F$ is the electromagnetic field tensor with dual $\tilde{F}$, and $g_{a\gamma}$ is the coupling.

The production of ALP coupling to photons is driven primarily by Primakoff conversion $N+\gamma\to N+a$ and coalescence $\gamma+\gamma\to a$; the latter dominates for heavy ALPs~\cite{Caputo:2022mah}. The volumetric emissivity from these processes depends on the temperature and density of the hot plasma, and is reviewed in the Supplemental Material (SM)~\cite{supplementalmaterial}. The overall amount of ALPs emitted by the SN core is obtained by integrating this emissivity over the profile of the SN core for the Garching group's muonic model SFHo-18.8~\cite{Bollig:2020xdr}. This model exhibits a relatively cold core, with a maximum temperature reaching up to 40~MeV, with a baryonic PNS mass of $1.351\, M_\odot$, and thus leads to a conservative estimate for the amount of ALPs that a typical core-collapse SN could produce; it is often referred to as cold SN model in the literature (e.g., Refs.~\cite{Caputo:2021rux,Fiorillo:2022cdq,Fiorillo:2024upk,Telalovic:2024cot}). Since we are interested in heavy ALPs, at the kinematic threshold for production and therefore influenced by inhomogeneous regions with high temperature, we do not attempt a reproduction of the resulting ALP emissivity using a one-zone homogeneous scenario. The final result is contained in the total amount of ALPs emitted during the core collapse $dN_a/dE_a$ differential in the ALP energy $E_a$, which we show in SM~\cite{supplementalmaterial}.

The ALPs produced in the core decay to gamma rays via $a\to \gamma+\gamma$. Those photons produced outside of the progenitor, with a radius $R_{\rm prog}$, will source a flash  visible at Earth; the resulting time-integrated photon spectrum is
\begin{equation}\label{eq:time_integrated_spectrum}
    \frac{dN_\gamma}{dE_\gamma}=\int_{E_\gamma+\frac{m_a^2}{4E_\gamma}}^{+\infty} \frac{2 dE_a}{p_a}\frac{dN_a}{dE_a}\exp\left[-\frac{\Gamma_a m_a R_{\rm prog}}{p_a}\right] F_{\rm sup};
\end{equation}
here $\Gamma_a=g_{a\gamma}^2m_a^3/64\pi$ is the decay rate of an ALP at rest with mass $m_a$, $p_a=\sqrt{E_a^2-m_a^2}$ is the axion momentum; see, e.g., Ref.~\cite{Fiorillo:2025sln} for a derivation. The signal depends on the coupling and mass. We also introduce a suppression factor due to a cap on the time window $t_{\rm max}$,
\begin{equation}
    F_{\rm sup}=1-\exp\left[-\Gamma_a\left(\frac{2E_\gamma t_{\rm max}}{m_a}-\frac{m_a R_{\rm prog}}{p_a}\right)\right].
\end{equation}
From here we see that the signal is spread over a typical time span determined by the propagation time for photons produced from noncollinear decays and for the slower-than-light ALPs before the decay~\cite{Telalovic:2024cot}
\begin{equation}
    t_{\rm del}=\frac{m_a}{2\Gamma_a E_a}\simeq 220\,\mathrm{s}\,g_{10}^{-2}E_{300}^{-1}m_{10}^{-2},
\end{equation}
with $g_{10}=g_{a\gamma}/10^{-10}\,\mathrm{GeV}^{-1}$, $E_{300}=E_a/300\,\mathrm{MeV}$, and $m_{10}=m_a/10\,\mathrm{MeV}$. When $t_{\rm del}\lesssim 5\,\mathrm{s}$, the typical duration of the SN emission of feebly interacting particles (see, e.g., Ref.~\cite{Fiorillo:2023frv} for the case of neutrinos), the duration of the signal will be determined by the SN emission itself. For our purposes, as we will see, it suffices that the flash lasts less than an hour or so to treat it as an instantaneous pulse, for which only the time-integrated emission in Eq.~\eqref{eq:time_integrated_spectrum} matters. Hence, we choose $t_{\rm max}$ equal to 1~h, the span of one bin, discussing the implications below.

The advantage of Type Ic SNe, compared with SN~1987A, appears now clear; the smaller $R_{\rm prog}$, the larger couplings can be probed without the exponential decay term in Eq.~\eqref{eq:time_integrated_spectrum} suppressing the signal. For a typical Type Ic SN, the progenitor radius used throughout this Letter is $R_{\rm prog}=10^{11}\,\mathrm{cm}$. The typical radii of C+O/He composition interfaces in red supergiants are much lower, ranging from 3--$6\times 10^9\, \mathrm{cm}$ (e.g., Ref.~\cite{Wongwathanarat:2014yda}). For naked C+O stars, observational constraints are harder to come by. Simulations of helium stars including mass loss lead to typical radii of $10^{10-11}\,\mathrm{cm}$~\cite{Yoon:2010at,2019ApJ...878...49W}, but Type Ic are for the larger part expected to have lost the helium envelope as well. Nevertheless, since we will consider a sample of Ic SNe, we cannot exclude that some of them may have some residual helium.  Ultrastripped progenitors may sport radii as low as $10^{10}\,\mathrm{cm}$~\cite{Tauris:2015xra}. For individual Type Ibc SNe, the radii obtained from light-curve fitting can be $10^{10}\,\mathrm{cm}$~\cite{1994ApJ...437L.115I,Woosley:1998hk}. For comparison, the radius of Sanduleak, the progenitor of SN~1987A, had a radius of about $35\,R_\odot\simeq 2.5\times 10^{12}\,\mathrm{cm}$, more than 2 orders of magnitude larger \cite{Menon+2019,Ono+2020,Orlando+2020,Utrobin+2021}. Since the decay rate grows with the square of the axion-photon coupling, by this simple argument we can probe couplings one order of magnitude larger than SN~1987A, provided we identify potential sources from which a gamma-ray flash should have been visible.

\begin{figure}[t]
    \centering
    \includegraphics[width=1.\columnwidth]{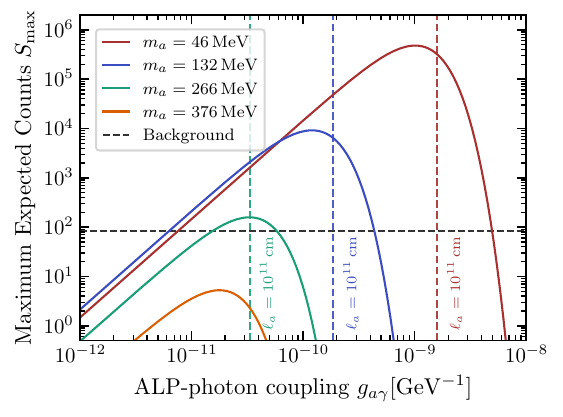}
    \caption{Expected signal counts for SN~2020oi as a function of $g_{a\gamma}$ for different ALP masses. We obtain the expected counts from the time-integrated model fluence, assuming the maximum exposure among all the 1-h time bins within the time window of interest. The horizontal line shows the mean expected background counts per 1-h bin in the time window $v_1$. The vertical lines correspond to a decay length $\ell_a$ for an ALP with energy $300$~MeV equal to the progenitor radius.}
    \label{fig:expected_events}
\end{figure}

{\bf\textit{Fermi-LAT detection of gamma-ray flash}}---If the decay length were long enough, the photons produced in the decay of ALPs would propagate freely, arriving at Earth and potentially showing up at gamma-ray telescopes. Here we focus on a collection of SNe, motivated by their close distance and the possibility of pinpointing rather precisely their explosion time $t_{\mathrm{coll}}$; all sources are collected in SM (Table~I)~\cite{supplementalmaterial}, together with their distance $d$ from Earth. The corresponding differential fluence expected at Earth is then $\mathcal{F}_\gamma=\frac{1}{4\pi d^2}dN_\gamma/dE_\gamma$.

Barring fireball formation (see below), ALPs coupling to photons should produce a gamma-ray flash at 100-MeV energies, i.e., at energies much larger than the usually observed energies of gamma-ray bursts (GRBs), for all Type Ic SNe. We therefore ask whether such a signal should have been observable with the Fermi Large Area Telescope (LAT), which is sensitive to $\gamma$-rays from 20 MeV to beyond 300 GeV \cite{Fermi-LAT:2009ihh}.

At first glance, constraining gamma-ray flashes from Type Ic SNe with Fermi-LAT appears impossible due to a timing issue. The axion-induced flash lasts between a few seconds and a few hours, and its initial time is impossible to pinpoint in the absence of a visible neutrino burst due to the large distance from the source. While a candidate SN can be precisely identified, the beginning of the explosion is usually uncertain; the stellar core collapse happens much earlier and the explosion becomes observable with a delay determined by the propagation time of the shock wave through the progenitor, which is typically less than a minute for the considered compact progenitors. The difficulty in estimating the time of arrival of the putative fireball implies that we cannot even know if Fermi-LAT, which only sees a portion of the sky of about 20\% at any given time, and whose field of view
varies over timescales of minutes, was actually sensitive during the collapse. Indeed, if a clear timing of the onset of the explosion is not available, a single Type Ic SN cannot individually provide constraints; it is always possible that the explosion occurred while Fermi-LAT was not observing the correct direction. (Notice that this statement does not necessarily hold for a broad-lined Type Ic SN associated with a detected high-energy burst, for which $t_{\rm coll}$ could be constrained accurately.)
However, the relatively frequent occurrence of Type Ic SNe allows us to circumvent this difficulty, since the probability of a missed observation for all the SNe within a sample is significantly suppressed.

Motivated by this simple argument, we consider a sample of 15 Type Ic SNe, and consider their expected signal at Fermi-LAT. We use time bins of 1~h, consistently with our choice of $t_{\rm max}$, and neglect any portion of the signal beyond that limit. We comment on the impact of this choice in our discussion.
The details of the Fermi-LAT analysis we perform are reported in SM~\cite{supplementalmaterial}.

In Fig.~\ref{fig:expected_events}, we show the number of signal events for one of our SNe (SN~2020oi), as a function of the coupling $g_{a\gamma}$, for various ALP masses. At low couplings, the expected signal grows quadratically with the coupling, while at large couplings it drops exponentially, due to a large fraction of ALPs decaying within the mantle of the progenitor. The threshold coupling at which exponential suppression begins is determined by the approximate condition $\Gamma_a R_{\rm prog}\gtrsim 1$ [see Eq.~\eqref{eq:time_integrated_spectrum}], as we can clearly see from the vertical lines in Fig.~\ref{fig:expected_events}; since $R_{\rm prog}$ is much lower for our Type Ic SNe compared with Type II SNe, the signal remains large at much higher couplings than SN~1987A and other candidate SNe.

To finally constrain the ALP-photon coupling, we consider for each SN (indexed by Latin letters $i$) in our sample the probability that Fermi-LAT would not observe any gamma-ray flash within a $\pm 3$-day window around $t_{\rm coll}$. This window was selected based on the uncertainty on $t_{\rm coll}$, as discussed in SM~\cite{supplementalmaterial}. Using the reconstructed uncertainty, we determine the probability that the flash happened in each temporal bin (indexed by Greek letters $\alpha$)  $P_{i,\alpha}$, the number of background events $B_{i,\alpha}$, the observed number of events $N_{i,\alpha}$, and the expected number of events $S_{i,\alpha}$. For many of the bins the experiment is blind, so $S_{i,\alpha}=0$.

We assume a Poisson likelihood for the number count $N_{i,\alpha}$, and denote the overall likelihood, including all bins and SNe, as $\Pi(g_{a\gamma},m_a)$, with the renormalized log-likelihood $\Lambda(g_{a\gamma},m_a)=-2\log[\Pi/\Pi^0]$, where $\Pi^0$ is the likelihood in the absence of a gamma-ray flash. We then define a test statistic as the log-likelihood ratio with the best-fit case
\begin{equation}
    \chi(g_{a\gamma},m_a)=
    \Lambda(g_{a\gamma},m_a)-\Lambda(\hat{g}_{a\gamma}(m_a),m_a),
\end{equation}
where $\hat{g}_{a\gamma}(m_a)$ is the ALP-photon coupling maximizing the likelihood for each value of $m_a$. For a $90\%$ confidence level we set constraints by requiring $\chi(g_{a\gamma},m_a)=2.71$. With this strategy, even if a single SN has a large probability $f_i$ for the gamma-ray flash to be missed, for $N$ SNe this probability drops exponentially with $N$, allowing to set constraints.

Figure~\ref{fig:results_exclusion} shows our final constraints on the ALP-photon coupling in the small-coupling regime, marked by ``stacked events.'' As expected, these constraints reach decay lengths shorter than the ones excluded by SN~1987A because the more compact progenitor lets ALPs escape more easily.

\begin{figure}
    \centering
    \includegraphics[width=0.5\textwidth]{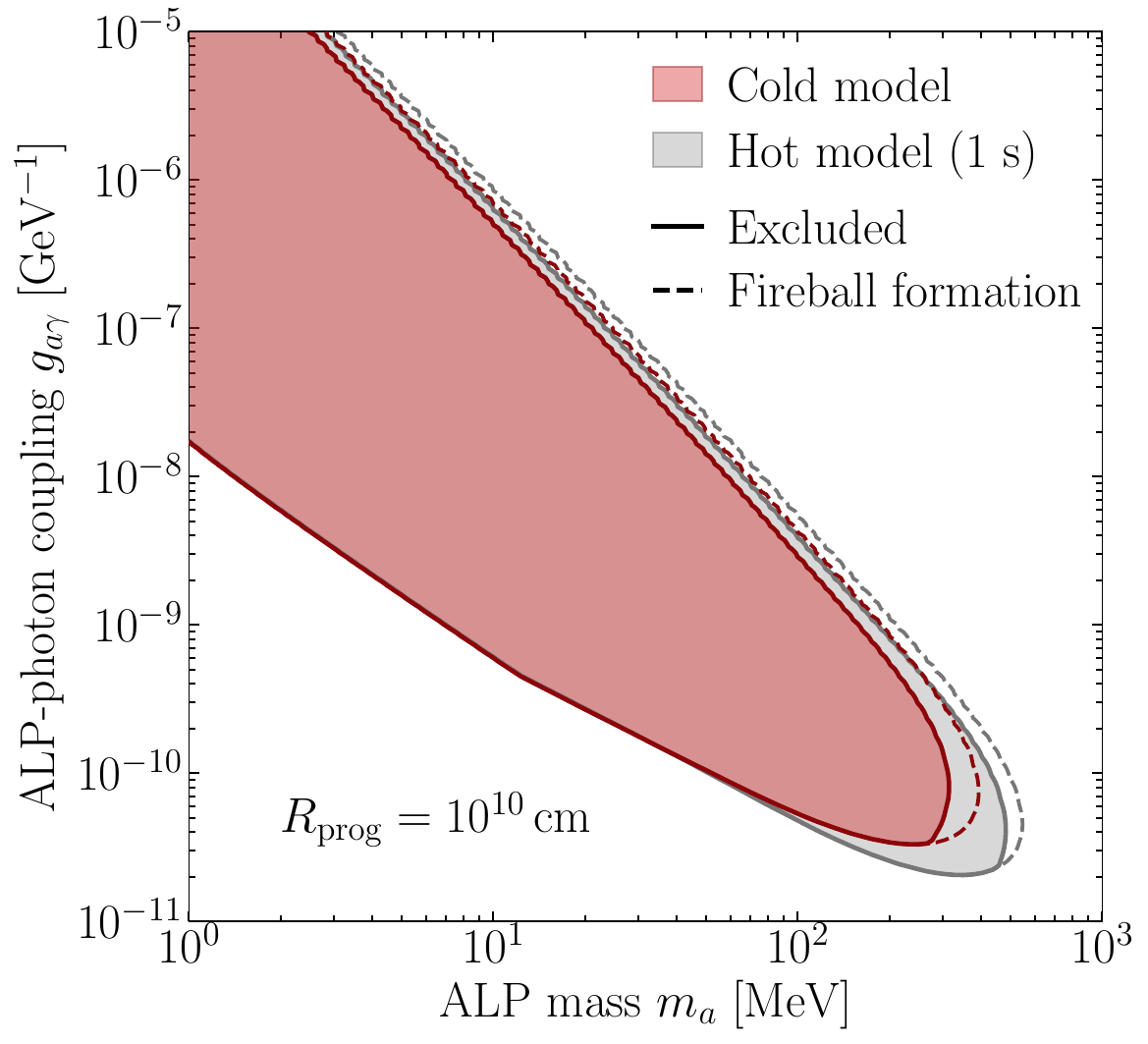}
    \caption{Region of fireball formation (dashed) and constrained region from fireball nonobservation (solid; shaded). Our fiducial model for ALP production is the cold SN Garching simulation (red); we also show results using a hot SN model (gray) capped at 1~s postbounce, coarsely modeling a black hole formation in the broad-lined SN~1998bw.}
    \label{fig:fireball_constraints}
\end{figure}

{\bf\textit{ALP-sourced fireball at SN~1998bw}}---A peculiarity of the region we are constraining is that, if photons are injected in a compact enough region, they produce  electron-positron pairs that subsequently lower their energy by radiating more photons via bremsstrahlung. The gamma rays, originally at hundreds of MeV, are reprocessed into a \textit{fireball}~\cite{Diamond:2023scc}, and their final energy ends up in the hundreds of keV range. (Similar fireballs can also be formed for strongly self-interacting neutrinos~\cite{Fiorillo:2023cas,Fiorillo:2023ytr} and dark particles~\cite{Fiorillo:2024upk}.) The region of fireball formation is deduced following the procedure in Refs.~\cite{Diamond:2023scc, Diamond:2023cto}, which we recollect in SM~\cite{supplementalmaterial}, and is shown in Fig.~\ref{fig:fireball_constraints}; in this region, Fermi-LAT constraints do not apply, and must be complemented by an analysis of the x-ray signal from the direction of the relevant SNe. We show the fireball region for two alternative models of the SN core: the cold SN model, which is meant to represent qualitatively a scenario with formation of a rotating protomagnetar~\cite{Mazzali+2014,Aloy+2021}, and the so-called hot SN model LS220-20.0 with a typically higher temperature in the core~\cite{Mirizzi+2016,Bollig:2020xdr} cut off at 1~s postbounce, meant to represent the more typical scenario for broad-lined SN in which the jet launching follows the formation of a black hole that shuts down emission from the PNS~\cite{Woosley1993,Woosley+2006}. For our fiducial bounds in Fig.~\ref{fig:results_exclusion}, we conservatively use the former scenario.

Crucially, the rare broad-lined Type~Ic SN~1998bw has famously been related to GRB 980425, resulting in the first corroboration of the link between GRBs and SNe~\cite{Kulkarni+1998,Cano:2016ccp}. The GRB was observed by the Burst and Transient Source Experiment (BATSE) instrument on the Compton Gamma-Ray Observatory~\cite{1998GCN....67....1K} and by BeppoSAX~\cite{Galama:1998ea}, as both experiments were online on April 25, 1998, at 21:49 UTC. Differently from the SNe considered in the Fermi-LAT analysis, we can constrain the time of collapse quite accurately, thanks to the burst produced by the jet launched by the spinning protomagnetar or the black hole eventually formed after the collapse.  The estimated distance was $\sim 40\,\rm Mpc$~\cite{1998Natur.395..670G,Soderberg:2004ht,Mirabal:2022uml}; the measured fluence integrated above 20~keV from BATSE was $(4.0\pm 0.6)\times 10^{-6}\,\mathrm{erg/cm}^2$~\cite{1998GCN....67....1K}, so that a limit on the gamma-ray flash from ALP decay is $7\times 10^{47}$~erg. This allows us to set constraints on the nonobservation of the reprocessed gamma-ray signal in the fireball region, constraining the total energy injected in ALPs. Notice that we are extremely conservative through this criterion, since in reality no gamma rays were observed above 300~keV, and the constrained fluence refers to a duration of 23~s, while our ALP signal would be spread over roughly 1~s. The energy range of BATSE coincides with the range in which the fireball spectrum is reprocessed~\cite{Diamond:2023cto}. The overall region constrained by the fireball argument is shown in Fig.~\ref{fig:results_exclusion}.
We finally stress that Fig.~\ref{fig:fireball_constraints} is obtained for $R_{\rm prog}=10^{10}\,\mathrm{cm}$, which was obtained for the specific case of SN~1998bw from light-curve fitting~\cite{Woosley:1998hk}, while the stacked-events bound assumes $R_{\rm prog}=10^{11}\,\mathrm{cm}$. Therefore, at least part of the fireball formation region in Fig.~\ref{fig:fireball_constraints} can be excluded also with the Fermi-LAT analysis.

{\bf\textit{Results and outlook}}---We have provided the first proof-of-principle use of Type~Ic SNe in constraining new radiatively decaying particles by relying on two different observables: the direct observation of gamma rays produced in the decay (small-coupling regime), and the observation of ALP-sourced fireballs (large-coupling regime). The new perspectives offered by Type~Ic SNe arise from their compact progenitors, less shielding on particles with short decay length. By a similar token, these compact progenitors host stronger magnetic fields with a new target of opportunity for axion-photon conversions~\cite{Candon:2025sdm}.

The stacked-events strategy is efficient at relatively small couplings, yet still reaches above the SN~1987A bounds at large masses. Our choice of integrating the signal only within $t_{\rm max}=1~\mathrm{h}$ is conservative, but most importantly inessential at the large couplings where our constraints supersede the SN~1987A. The typical delay overshoots $t_{\rm max}$ only for very small couplings, corresponding to the kink in the constraint in Fig.~\ref{fig:results_exclusion} around $m_a\sim 130\,\mathrm{MeV}$, a region largely excluded by SN~1987A.

Part of the fireball region overlaps with the region excluded from the neutron star merger GW170817~\cite{Diamond:2023cto}. On the other hand, the latter carries strong intrinsic uncertainty due to the potential presence of ejected material from the merger that could trap the photons from decay; its impact on the GW170817 bounds is visible in the Supplemental Material of Ref.~\cite{Diamond:2023cto}. Hence, our new bounds, which are entirely insensitive to this systematic uncertainty, are a powerful addition to the constraining strategies on decaying particles.

Next-generation $\gamma$-ray instruments with $4\pi$ continuous sky coverage, such as the proposed constellation of SmallSats GALAXIS in the energy range $100$--$500\,\mathrm{MeV}$ \cite{Manzari:2024jns}, could strengthen our constraints substantially by removing the temporal dead times inherent to scanning single telescopes such as Fermi-LAT.  Figure~\ref{fig:results_exclusion} shows a projected exclusion for a future Type~Ic SN at a distance of $10\,\mathrm{Mpc}$---multiple SNe in our samples lie at comparable distances---with progenitor radius $R_{\rm prog}=10^{10}\,\mathrm{cm}$.
We use the highest exposure time bin in the SN2020oi signal window, and set bounds from the Poisson likelihood for a nondetection. Hence, an extragalactic Type Ic SN, coupled with a gamma-ray detector with $4\pi$ coverage, would surpass the SN~1987A bounds at all ALP masses. When considering extragalactic distances, Type Ic SNe have a large occurrence rate: core-collapse SNe account for nearly 70\% of all SNe, and of the core-collapse SNe 23\% are SNe Ibc. In turn, Type Ic SNe constitute around 40\% of SNe~Ibc (60\% when including broad-line SNe~Ic-BL)~\cite{Ma:2025css}. Therefore, using the local core-collapse SN rate from~\cite{Lecce:2025vjc}, the SN~Ic rate is around 0.2 per year up to $10\,\rm Mpc$, and around 0.7 per year up to $20\,\rm Mpc$ (including SNe~Ic-BL these numbers are 0.3 and 1.05 per year, respectively).

The new observables follow the track of a general collection of new strategies to observe or constrain sub-GeV particles, based either on their energy deposition in stellar progenitors~\cite{Falk:1978kf,Sung:2019xie,Caputo:2022mah,Fiorillo:2025yzf} or outside of them, with the production of visible messengers (neutrinos~\cite{Mastrototaro:2019vug,Fiorillo:2022cdq,Akita:2022etk,Telalovic:2024cot} or photons, either directly~\cite{Chupp:1989kx,Kolb:1988pe,Oberauer:1993yr,Jaffe:1995sw,Jaeckel:2017tud,Caputo:2021rux,Hoof:2022xbe,Candon:2024eah,Benabou:2024jlj,Chen:2024ekh} or through a reprocessing due to fireball formation~\cite{Diamond:2023scc,Diamond:2023cto}). Such observables extend far beyond the simple ALP-photon coupling; while we use this case as a benchmark, analogous bounds can be drawn on any model with particles decaying to leptonic or hadronic sectors, such as ALPs coupling to electrons---see, e.g., the recent compilation of constraints in Ref.~\cite{Fiorillo:2025sln}---or sterile neutrinos~\cite{Carenza:2023old}.

\textbf{\textit{Acknowledgments}}---We warmly thank Francesca Calore for her valuable comments and guidance on the time-dependent Fermi-LAT analysis applied in this work. F. R. C. acknowledges support from the Deutsche Forschungsgemeinschaft (DFG, German Research Foundation) under Germany's Excellence Strategy---Cluster of Excellence ``Color meets Flavor,'' EXC 3107---Project-ID 533766364, and from the German federal and state program ``Professorinnenprogramm 2030'' Project-ID 01FP24167Q. D. F. G. F. is supported by the Alexander von Humboldt Foundation (Germany). H. T. J. acknowledges support by the German Research Foundation (DFG) through the Collaborative Research Centre ``Neutrinos and Dark Matter in Astro- and Particle Physics (NDM),'' Grant SFB-1258-283604770, and under Germany's Excellence Strategy through the Cluster of Excellence ORIGINS EXC-2094-390783311. B. v. B. is supported by the European Research Council (ERC) under the European Union's Horizon 2020 Research and Innovation Programme (ERC Starting Grant No.~[803189]). E. V. is supported by the Italian MUR Departments of
Excellence grant 2023--2027 ``Quantum Frontier,'' by the Italian MUR through the FIS 2 project FIS-2023-01577 (DD no.~23314 December 10, 2024, CUP C53C24001460001), and by Istituto Nazionale di Fisica Nucleare (INFN) through the Theoretical Astroparticle Physics (TAsP) project.

\textbf{\textit{Data availability}}---
The Fermi-LAT data used in this work are publicly available from the Fermi Science Support Center. The supernova sample used in the analysis is compiled in Table~I of Supplemental Material~\cite{supplementalmaterial}. The numerical data and custom scripts used to generate the figures and constraints are available from the authors upon reasonable request.

\bibliographystyle{bibi}
\bibliography{References}

@ARTICLE{Kulkarni+1998,
       author = {{Kulkarni}, S.~R. and {Frail}, D.~A. and {Wieringa}, M.~H. and {Ekers}, R.~D. and {Sadler}, E.~M. and {Wark}, R.~M. and {Higdon}, J.~L. and {Phinney}, E.~S. and {Bloom}, J.~S.},
        title = "{Radio emission from the unusual supernova 1998bw and its association with the {\ensuremath{\gamma}}-ray burst of 25 April 1998}",
      journal = {Nature},
         year = 1998,
        month = oct,
       volume = {395},
       number = {6703},
        pages = {663-669},
          doi = {10.1038/27139},
       adsurl = {https://ui.adsabs.harvard.edu/abs/1998Natur.395..663K}
}

@ARTICLE{Woosley1993,
       author = {{Woosley}, S.~E.},
        title = "{Gamma-Ray Bursts from Stellar Mass Accretion Disks around Black Holes}",
      journal = {Astrophys. J.},
         year = 1993,
        month = mar,
       volume = {405},
        pages = {273},
          doi = {10.1086/172359},
       adsurl = {https://ui.adsabs.harvard.edu/abs/1993ApJ...405..273W}
}

@ARTICLE{Woosley+2006,
       author = {{Woosley}, S.~E. and {Bloom}, J.~S.},
        title = "{The Supernova Gamma-Ray Burst Connection}",
      journal = {Ann. Rev. Astron. \& Astrophys.},
         year = 2006,
        month = sep,
       volume = {44},
       number = {1},
        pages = {507-556},
          doi = {10.1146/annurev.astro.43.072103.150558},
archivePrefix = {arXiv},
       eprint = {astro-ph/0609142},
 primaryClass = {astro-ph},
       adsurl = {https://ui.adsabs.harvard.edu/abs/2006ARA&A..44..507W}
}

@ARTICLE{Mazzali+2014,
       author = {{Mazzali}, P.~A. and {McFadyen}, A.~I. and {Woosley}, S.~E. and {Pian}, E. and {Tanaka}, M.},
        title = "{An upper limit to the energy of gamma-ray bursts indicates that GRBs/SNe are powered by magnetars}",
      journal = {Mon. Not. Roy. Astron. Soc.},
         year = 2014,
        month = sep,
       volume = {443},
       number = {1},
        pages = {67-71},
          doi = {10.1093/mnras/stu1124},
archivePrefix = {arXiv},
       eprint = {1406.1209},
 primaryClass = {astro-ph.HE},
       adsurl = {https://ui.adsabs.harvard.edu/abs/2014MNRAS.443...67M}
}

@ARTICLE{Aloy+2021,
       author = {{Aloy}, M. {\'A}. and {Obergaulinger}, M.},
        title = "{Magnetorotational core collapse of possible GRB progenitors - II. Formation of protomagnetars and collapsars}",
      journal = {Mon. Not. Roy. Astron. Soc.},
         year = 2021,
        month = jan,
       volume = {500},
       number = {4},
        pages = {4365-4397},
          doi = {10.1093/mnras/staa3273},
archivePrefix = {arXiv},
       eprint = {2008.03779},
 primaryClass = {astro-ph.HE},
       adsurl = {https://ui.adsabs.harvard.edu/abs/2021MNRAS.500.4365A}
}

@ARTICLE{Mirizzi+2016,
       author = {{Mirizzi}, A. and {Tamborra}, I. and {Janka}, H. -Th. and {Saviano}, N. and {Scholberg}, K. and {Bollig}, R. and {H{\"u}depohl}, L. and {Chakraborty}, S.},
        title = "{Supernova neutrinos: production, oscillations and detection}",
      journal = {Nuovo Cimento Rivista Serie},
         year = 2016,
        month = feb,
       volume = {39},
       number = {1-2},
        pages = {1-112},
          doi = {10.1393/ncr/i2016-10120-8},
archivePrefix = {arXiv},
       eprint = {1508.00785},
 primaryClass = {astro-ph.HE},
       adsurl = {https://ui.adsabs.harvard.edu/abs/2016NCimR..39....1M}
}

@misc{supplementalmaterial,
  note = {See the Supplemental Material for a discussion of the
  analogy between decay bounds from SNe and beam-dump experiments,
  the production and subsequent decay of ALPs from SNe, including the
  potential formation of a fireball for sufficiently large couplings,
  additional information on the SNe used in this work, and the
  statistical analysis adopted, which includes
  Refs.~\cite{
  Ghosh:2025rjh,
  Goodman:1986az,
  zhang2019physics,
  Fiorillo:2025gnd,
  Diamond:2021ekg,
  Barbarino:2020amq,
  Meyer:2020vzy,
  Tinyanont:2016vbz,
  Barbarino:2017mes,
  Prentice:2018ual,
  Kankare:2021fzf,
  YoungSupernovaExperiment:2021fur,
  Muller:2023pip,
  Muller:2023vjm,
  fermitools,
  Wilks:1938dza}.},
}

@misc{fermitools,
  note = {Reference page for Fermitools: \href{https://fermi.gsfc.nasa.gov/ssc/data/analysis/software/}{https://fermi.gsfc.nasa.gov/ssc/data/analysis/software/}} 
}

@ARTICLE{1998GCN....67....1K,
       author = {{Kippen}, R.~M. and {BATSE Grb Team}},
        title = "{GRB980425 BATSE observations}",
      journal = {GRB Coordinates Network},
         year = 1998,
        month = may,
       volume = {67},
        pages = {1},
       adsurl = {https://ui.adsabs.harvard.edu/abs/1998GCN....67....1K}
}

@article{Cano:2016ccp,
    author = "Cano, Zach and Wang, Shan-Qin and Dai, Zi-Gao and Wu, Xue-Feng",
    title = "{The Observer{\textquoteright}s Guide to the Gamma-Ray Burst Supernova Connection}",
    eprint = "1604.03549",
    archivePrefix = "arXiv",
    primaryClass = "astro-ph.HE",
    doi = "10.1155/2017/8929054",
    journal = "Adv. Astron.",
    volume = "2017",
    pages = "8929054",
    year = "2017"
}

@article{Galama:1998ea,
    author = "Galama, T. J. and others",
    title = "{Discovery of the peculiar supernova 1998bw in the error box of GRB 980425}",
    eprint = "astro-ph/9806175",
    archivePrefix = "arXiv",
    doi = "10.1038/27150",
    journal = "Nature",
    volume = "395",
    pages = "670",
    year = "1998"
}

@article{Candon:2025sdm,
    author = "Cand{\'o}n, Francisco R. and Fiorillo, Damiano F. G. and Gil Muyor, {\'A}ngel and Janka, Hans-Thomas and Raffelt, Georg G. and Vitagliano, Edoardo",
    title = "{Stripped-Envelope Supernovae for QCD Axion Detection}",
    eprint = "2511.13815",
    archivePrefix = "arXiv",
    primaryClass = "hep-ph",
    month = "11",
    year = "2025"
}

@ARTICLE{Menon+2019,
       author = {{Menon}, Athira and {Utrobin}, Victor and {Heger}, Alexander},
        title = "{Explosions of blue supergiants from binary mergers for SN 1987A}",
      journal = {Mon. Not. Roy. Astron. Soc.},
         year = 2019,
        month = jan,
       volume = {482},
       number = {1},
        pages = {438-452},
          doi = {10.1093/mnras/sty2647},
archivePrefix = {arXiv},
       eprint = {1806.08072},
 primaryClass = {astro-ph.SR},
       adsurl = {https://ui.adsabs.harvard.edu/abs/2019MNRAS.482..438M}
}

@ARTICLE{Utrobin+2021,
       author = {{Utrobin}, V.~P. and {Wongwathanarat}, A. and {Janka}, H. -Th. and {M{\"u}ller}, E. and {Ertl}, T. and {Menon}, A. and {Heger}, A.},
        title = "{Supernova 1987A: 3D Mixing and Light Curves for Explosion Models Based on Binary-merger Progenitors}",
      journal = {Astrophys. J.},
         year = 2021,
        month = jun,
       volume = {914},
       number = {1},
          eid = {4},
        pages = {4},
          doi = {10.3847/1538-4357/abf4c5},
archivePrefix = {arXiv},
       eprint = {2102.09686},
 primaryClass = {astro-ph.HE},
       adsurl = {https://ui.adsabs.harvard.edu/abs/2021ApJ...914....4U}
}

@ARTICLE{Ono+2020,
       author = {{Ono}, Masaomi and {Nagataki}, Shigehiro and {Ferrand}, Gilles and {Takahashi}, Koh and {Umeda}, Hideyuki and {Yoshida}, Takashi and {Orlando}, Salvatore and {Miceli}, Marco},
        title = "{Matter Mixing in Aspherical Core-collapse Supernovae: Three-dimensional Simulations with Single-star and Binary Merger Progenitor Models for SN 1987A}",
      journal = {Astrophys. J.},
         year = 2020,
        month = jan,
       volume = {888},
       number = {2},
          eid = {111},
        pages = {111},
          doi = {10.3847/1538-4357/ab5dba},
archivePrefix = {arXiv},
       eprint = {1912.02234},
 primaryClass = {astro-ph.HE},
       adsurl = {https://ui.adsabs.harvard.edu/abs/2020ApJ...888..111O}
}

@ARTICLE{Orlando+2020,
       author = {{Orlando}, S. and {Ono}, M. and {Nagataki}, S. and {Miceli}, M. and {Umeda}, H. and {Ferrand}, G. and {Bocchino}, F. and {Petruk}, O. and {Peres}, G. and {Takahashi}, K. and {Yoshida}, T.},
        title = "{Hydrodynamic simulations unravel the progenitor-supernova-remnant connection in SN 1987A}",
      journal = {Astron. Astrophys.},
         year = 2020,
        month = apr,
       volume = {636},
          eid = {A22},
        pages = {A22},
          doi = {10.1051/0004-6361/201936718},
archivePrefix = {arXiv},
       eprint = {1912.03070},
 primaryClass = {astro-ph.HE},
       adsurl = {https://ui.adsabs.harvard.edu/abs/2020A&A...636A..22O}
}

@inproceedings{Essig:2013lka,
    author = "Essig, Rouven and others",
    title = "{Working Group Report: New Light Weakly Coupled Particles}",
    booktitle = "{Snowmass 2013}: {Snowmass on the Mississippi}",
    eprint = "1311.0029",
    archivePrefix = "arXiv",
    primaryClass = "hep-ph",
    reportNumber = "YITP-SB-36, FERMILAB-CONF-13-653",
    month = "10",
    year = "2013"
}

@book{Raffelt:1996wa,
    author = "Raffelt, G. G.",
    title = "{Stars as laboratories for fundamental physics}: {The astrophysics of neutrinos, axions, and other weakly interacting particles}",
    isbn = "978-0-226-70272-8",
    month = "5",
    year = "1996",
    publisher= "University of Chicago Press, Chicago"
}

@article{Kolb:1988pe,
    author = "Kolb, Edward W. and Turner, Michael S.",
    title = "{Limits to the Radiative Decays of Neutrinos and Axions from Gamma-Ray Observations of SN 1987a}",
    reportNumber = "FERMILAB-PUB-87-223-A, FERMILAB-PUB-87-223-A-REV",
    doi = "10.1103/PhysRevLett.62.509",
    journal = "Phys. Rev. Lett.",
    volume = "62",
    pages = "509",
    year = "1989"
}

@article{DeRocco:2019njg,
    author = "DeRocco, William and Graham, Peter W. and Kasen, Daniel and Marques-Tavares, Gustavo and Rajendran, Surjeet",
    title = "{Observable signatures of dark photons from supernovae}",
    eprint = "1901.08596",
    archivePrefix = "arXiv",
    primaryClass = "hep-ph",
    doi = "10.1007/JHEP02(2019)171",
    journal = "JHEP",
    volume = "02",
    pages = "171",
    year = "2019"
}

@article{Jaffe:1995sw,
    author = "Jaffe, Andrew H. and Turner, Michael S.",
    title = "{Gamma-rays and the decay of neutrinos from SN1987A}",
    eprint = "astro-ph/9601104",
    archivePrefix = "arXiv",
    reportNumber = "FERMILAB-PUB-95-397-A, CITA-95-26",
    doi = "10.1103/PhysRevD.55.7951",
    journal = "Phys. Rev. D",
    volume = "55",
    pages = "7951--7959",
    year = "1997"
}

@article{Tauris:2015xra,
    author = "Tauris, Thomas M. and Langer, Norbert and Podsiadlowski, Philipp",
    title = "{Ultra-stripped supernovae: progenitors and fate}",
    eprint = "1505.00270",
    archivePrefix = "arXiv",
    primaryClass = "astro-ph.SR",
    doi = "10.1093/mnras/stv990",
    journal = "Mon. Not. Roy. Astron. Soc.",
    volume = "451",
    number = "2",
    pages = "2123--2144",
    year = "2015"
}

@article{Wongwathanarat:2014yda,
    author = {Wongwathanarat, Annop and M{\"u}ller, Ewald and Janka, H. -Thomas},
    title = "{Three-Dimensional Simulations of Core-Collapse Supernovae: From Shock Revival to Shock Breakout}",
    eprint = "1409.5431",
    archivePrefix = "arXiv",
    primaryClass = "astro-ph.HE",
    doi = "10.1051/0004-6361/201425025",
    journal = "Astron. Astrophys.",
    volume = "577",
    pages = "A48",
    year = "2015"
}

@article{Chupp:1989kx,
    author = "Chupp, E. L. and Vestrand, W. T. and Reppin, C.",
    title = "{Experimental Limits on the Radiative Decay of {SN1987A} Neutrinos}",
    doi = "10.1103/PhysRevLett.62.505",
    journal = "Phys. Rev. Lett.",
    volume = "62",
    pages = "505--508",
    year = "1989"
}

@article{Agrawal:2021dbo,
    author = "Agrawal, Prateek and others",
    title = "{Feebly-interacting particles: FIPs 2020 workshop report}",
    eprint = "2102.12143",
    archivePrefix = "arXiv",
    primaryClass = "hep-ph",
    doi = "10.1140/epjc/s10052-021-09703-7",
    journal = "Eur. Phys. J. C",
    volume = "81",
    number = "11",
    pages = "1015",
    year = "2021"
}

@article{Abdullahi:2022jlv,
    author = "Abdullahi, Asli M. and others",
    title = "{The present and future status of heavy neutral leptons}",
    eprint = "2203.08039",
    archivePrefix = "arXiv",
    primaryClass = "hep-ph",
    reportNumber = "FERMILAB-CONF-22-184-T-V",
    doi = "10.1088/1361-6471/ac98f9",
    journal = "J. Phys. G",
    volume = "50",
    number = "2",
    pages = "020501",
    year = "2023"
}

@article{Berryman:2022hds,
    author = "Berryman, Jeffrey M. and others",
    title = "{Neutrino self-interactions: A white paper}",
    note = "{(2022 Snowmass Summer Study)}",
    eprint = "2203.01955",
    archivePrefix = "arXiv",
    primaryClass = "hep-ph",
    reportNumber = "CERN-TH-2022-024, DESY-22-035, FERMILAB-PUB-22-099-T",
    doi = "10.1016/j.dark.2023.101267",
    journal = "Phys. Dark Univ.",
    volume = "42",
    pages = "101267",
    year = "2023"
}

@inproceedings{Baryakhtar:2022hbu,
    author = "Baryakhtar, Masha and others",
    title = "{Dark Matter In Extreme Astrophysical Environments}",
    booktitle = "{Snowmass 2021}",
    eprint = "2203.07984",
    archivePrefix = "arXiv",
    primaryClass = "hep-ph",
    month = "3",
    year = "2022"
}

@article{CHARM:1985anb,
    author = "Bergsma, F. and others",
    collaboration = "CHARM",
    title = "{Search for Axion Like Particle Production in 400-{GeV} Proton - Copper Interactions}",
    reportNumber = "CERN-EP-85-38",
    doi = "10.1016/0370-2693(85)90400-9",
    journal = "Phys. Lett. B",
    volume = "157",
    pages = "458--462",
    year = "1985"
}

@article{Knapen:2024fvh,
    author = "Knapen, Simon and Opferkuch, Toby and Redigolo, Diego and Tammaro, Michele",
    title = "{Displaced Searches for Axion-Like Particles and Heavy Neutral Leptons at Mu3e}",
    eprint = "2410.13941",
    archivePrefix = "arXiv",
    primaryClass = "hep-ph",
    month = "10",
    year = "2024"
}

@article{Krnjaic:2019rsv,
    author = "Krnjaic, Gordan and Marques-Tavares, Gustavo and Redigolo, Diego and Tobioka, Kohsaku",
    title = "{Probing Muonphilic Force Carriers and Dark Matter at Kaon Factories}",
    eprint = "1902.07715",
    archivePrefix = "arXiv",
    primaryClass = "hep-ph",
    reportNumber = "FERMILAB-PUB-18-665-A, KEK-TH-2105",
    doi = "10.1103/PhysRevLett.124.041802",
    journal = "Phys. Rev. Lett.",
    volume = "124",
    number = "4",
    pages = "041802",
    year = "2020"
}

@article{Calibbi:2020jvd,
    author = "Calibbi, Lorenzo and Redigolo, Diego and Ziegler, Robert and Zupan, Jure",
    title = "{Looking forward to lepton-flavor-violating ALPs}",
    eprint = "2006.04795",
    archivePrefix = "arXiv",
    primaryClass = "hep-ph",
    reportNumber = "P3H-20-024, TTP20-025",
    doi = "10.1007/JHEP09(2021)173",
    journal = "JHEP",
    volume = "09",
    pages = "173",
    year = "2021"
}

@article{Bauer:2017ris,
    author = "Bauer, Martin and Neubert, Matthias and Thamm, Andrea",
    title = "{Collider Probes of Axion-Like Particles}",
    eprint = "1708.00443",
    archivePrefix = "arXiv",
    primaryClass = "hep-ph",
    reportNumber = "MITP-17-047",
    doi = "10.1007/JHEP12(2017)044",
    journal = "JHEP",
    volume = "12",
    pages = "044",
    year = "2017"
}

@article{Bauer:2018onh,
    author = "Bauer, Martin and Foldenauer, Patrick and Jaeckel, Joerg",
    title = "{Hunting All the Hidden Photons}",
    eprint = "1803.05466",
    archivePrefix = "arXiv",
    primaryClass = "hep-ph",
    doi = "10.1007/JHEP07(2018)094",
    journal = "JHEP",
    volume = "07",
    pages = "094",
    year = "2018"
}

@article{Riordan:1987aw,
    author = "Riordan, E. M. and others",
    title = "{A Search for Short Lived Axions in an Electron Beam Dump Experiment}",
    reportNumber = "SLAC-PUB-4280, UR-993, FERMILAB-PUB-87-251",
    doi = "10.1103/PhysRevLett.59.755",
    journal = "Phys. Rev. Lett.",
    volume = "59",
    pages = "755",
    year = "1987"
}

@article{Blumlein:1990ay,
    author = "Blumlein, J. and others",
    title = "{Limits on neutral light scalar and pseudoscalar particles in a proton beam dump experiment}",
    reportNumber = "PHE-90-03",
    doi = "10.1007/BF01548556",
    journal = "Z. Phys. C",
    volume = "51",
    pages = "341--350",
    year = "1991"
}

@article{NA64:2020qwq,
    author = "Banerjee, D. and others",
    collaboration = "NA64",
    title = "{Search for Axionlike and Scalar Particles with the NA64 Experiment}",
    eprint = "2005.02710",
    archivePrefix = "arXiv",
    primaryClass = "hep-ex",
    reportNumber = "CERN-EP-2020-068",
    doi = "10.1103/PhysRevLett.125.081801",
    journal = "Phys. Rev. Lett.",
    volume = "125",
    number = "8",
    pages = "081801",
    year = "2020"
}

@article{Dolan:2017osp,
    author = "Dolan, Matthew J. and Ferber, Torben and Hearty, Christopher and Kahlhoefer, Felix and Schmidt-Hoberg, Kai",
    title = "{Revised constraints and Belle II sensitivity for visible and invisible axion-like particles}",
    eprint = "1709.00009",
    archivePrefix = "arXiv",
    primaryClass = "hep-ph",
    reportNumber = "DESY-17-127",
    doi = "10.1007/JHEP12(2017)094",
    journal = "JHEP",
    volume = "12",
    pages = "094",
    year = "2017",
    note = "[Erratum: JHEP 03, 190 (2021)]"
}

@article{Capozzi:2023ffu,
    author = "Capozzi, Francesco and Dutta, Bhaskar and Gurung, Gajendra and Jang, Wooyoung and Shoemaker, Ian M. and Thompson, Adrian and Yu, Jaehoon",
    title = "{New constraints on ALP couplings to electrons and photons from ArgoNeuT and the MiniBooNE beam dump}",
    eprint = "2307.03878",
    archivePrefix = "arXiv",
    primaryClass = "hep-ph",
    reportNumber = "MI-HET-808",
    doi = "10.1103/PhysRevD.108.075019",
    journal = "Phys. Rev. D",
    volume = "108",
    number = "7",
    pages = "075019",
    year = "2023"
}

@article{Magill:2018jla,
    author = "Magill, Gabriel and Plestid, Ryan and Pospelov, Maxim and Tsai, Yu-Dai",
    title = "{Dipole Portal to Heavy Neutral Leptons}",
    eprint = "1803.03262",
    archivePrefix = "arXiv",
    primaryClass = "hep-ph",
    reportNumber = "FERMILAB-PUB-18-745-A",
    doi = "10.1103/PhysRevD.98.115015",
    journal = "Phys. Rev. D",
    volume = "98",
    number = "11",
    pages = "115015",
    year = "2018"
}

@article{Bolton:2019pcu,
    author = "Bolton, Patrick D. and Deppisch, Frank F. and Bhupal Dev, P. S.",
    title = "{Neutrinoless double beta decay versus other probes of heavy sterile neutrinos}",
    eprint = "1912.03058",
    archivePrefix = "arXiv",
    primaryClass = "hep-ph",
    doi = "10.1007/JHEP03(2020)170",
    journal = "JHEP",
    volume = "03",
    pages = "170",
    year = "2020"
}

@article{Li:2025yzb,
    author = "Li, Haotian and Liu, Zuowei and Song, Ningqiang",
    title = "{Probing axion and muon-philic new physics with muon beam dump}",
    eprint = "2501.06294",
    archivePrefix = "arXiv",
    primaryClass = "hep-ph",
    month = "1",
    year = "2025"
}

@article{Oberauer:1993yr,
    author = "Oberauer, L. and Hagner, C. and Raffelt, G. and Rieger, E.",
    title = "{Supernova bounds on neutrino radiative decays}",
    doi = "10.1016/0927-6505(93)90004-W",
    journal = "Astropart. Phys.",
    volume = "1",
    pages = "377--386",
    year = "1993"
}

@article{Giannotti:2010ty,
    author = "Giannotti, M. and Duffy, L. D. and Nita, R.",
    title = "{New constraints for heavy axion-like particles from supernovae}",
    eprint = "1009.5714",
    archivePrefix = "arXiv",
    primaryClass = "astro-ph.HE",
    reportNumber = "LA-UR-10-05895",
    doi = "10.1088/1475-7516/2011/01/015",
    journal = "JCAP",
    volume = "01",
    pages = "015",
    year = "2011"
}

@article{Kazanas:2014mca,
    author = "Kazanas, Demos and Mohapatra, Rabindra N. and Nussinov, Shmuel and Teplitz, Vigdor L. and Zhang, Yongchao",
    title = "{Supernova Bounds on the Dark Photon Using its Electromagnetic Decay}",
    eprint = "1410.0221",
    archivePrefix = "arXiv",
    primaryClass = "hep-ph",
    reportNumber = "UMD-PP--014-015",
    doi = "10.1016/j.nuclphysb.2014.11.009",
    journal = "Nucl. Phys. B",
    volume = "890",
    pages = "17--29",
    year = "2014"
}

@article{Jaeckel:2017tud,
    author = "Jaeckel, J. and Malta, P. C. and Redondo, J.",
    title = "{Decay photons from the axionlike particles burst of type II supernovae}",
    eprint = "1702.02964",
    archivePrefix = "arXiv",
    primaryClass = "hep-ph",
    doi = "10.1103/PhysRevD.98.055032",
    journal = "Phys. Rev. D",
    volume = "98",
    number = "5",
    pages = "055032",
    year = "2018"
}

@article{Caputo:2021rux,
    author = "Caputo, Andrea and Raffelt, Georg and Vitagliano, Edoardo",
    title = "{Muonic boson limits: Supernova redux}",
    eprint = "2109.03244",
    archivePrefix = "arXiv",
    primaryClass = "hep-ph",
    reportNumber = "MPP-2021-154",
    doi = "10.1103/PhysRevD.105.035022",
    journal = "Phys. Rev. D",
    volume = "105",
    number = "3",
    pages = "035022",
    year = "2022"
}

@article{Caputo:2021kcv,
    author = "Caputo, Andrea and Carenza, Pierluca and Lucente, Giuseppe and Vitagliano, Edoardo and Giannotti, Maurizio and Kotake, Kei and Kuroda, Takami and Mirizzi, Alessandro",
    title = "{Axionlike Particles from Hypernovae}",
    eprint = "2104.05727",
    archivePrefix = "arXiv",
    primaryClass = "hep-ph",
    doi = "10.1103/PhysRevLett.127.181102",
    journal = "Phys. Rev. Lett.",
    volume = "127",
    number = "18",
    pages = "181102",
    year = "2021"
}

@article{Caputo:2022mah,
    author = "Caputo, Andrea and Janka, Hans-Thomas and Raffelt, Georg and Vitagliano, Edoardo",
    title = "{Low-Energy Supernovae Severely Constrain Radiative Particle Decays}",
    eprint = "2201.09890",
    archivePrefix = "arXiv",
    primaryClass = "astro-ph.HE",
    doi = "10.1103/PhysRevLett.128.221103",
    journal = "Phys. Rev. Lett.",
    volume = "128",
    number = "22",
    pages = "221103",
    year = "2022"
}

@article{Hoof:2022xbe,
    author = "Hoof, Sebastian and Schulz, Lena",
    title = "{Updated constraints on axion-like particles from temporal information in supernova SN1987A gamma-ray data}",
    eprint = "2212.09764",
    archivePrefix = "arXiv",
    primaryClass = "hep-ph",
    reportNumber = "TTP22-072",
    doi = "10.1088/1475-7516/2023/03/054",
    journal = "JCAP",
    volume = "03",
    pages = "054",
    year = "2023"
}

@article{Lella:2022uwi,
    author = "Lella, Alessandro and Carenza, Pierluca and Lucente, Giuseppe and Giannotti, Maurizio and Mirizzi, Alessandro",
    title = "{Protoneutron stars as cosmic factories for massive axionlike particles}",
    eprint = "2211.13760",
    archivePrefix = "arXiv",
    primaryClass = "hep-ph",
    doi = "10.1103/PhysRevD.107.103017",
    journal = "Phys. Rev. D",
    volume = "107",
    number = "10",
    pages = "103017",
    year = "2023"
}

@article{Fiorillo:2022cdq,
    author = "Fiorillo, Damiano F. G. and Raffelt, Georg G. and Vitagliano, Edoardo",
    title = "{Strong Supernova 1987A Constraints on Bosons Decaying to Neutrinos}",
    eprint = "2209.11773",
    archivePrefix = "arXiv",
    primaryClass = "hep-ph",
    doi = "10.1103/PhysRevLett.131.021001",
    journal = "Phys. Rev. Lett.",
    volume = "131",
    number = "2",
    pages = "021001",
    year = "2023"
}

@article{Akita:2022etk,
    author = "Akita, Kensuke and Im, Sang Hui and Masud, Mehedi",
    title = "{Probing non-standard neutrino interactions with a light boson from next galactic and diffuse supernova neutrinos}",
    eprint = "2206.06852",
    archivePrefix = "arXiv",
    primaryClass = "hep-ph",
    reportNumber = "CTPU-PTC-22-13",
    doi = "10.1007/JHEP12(2022)050",
    journal = "JHEP",
    volume = "12",
    pages = "050",
    year = "2022"
}

@article{Fiorillo:2023cas,
    author = "Fiorillo, Damiano F. G. and Raffelt, Georg G. and Vitagliano, Edoardo",
    title = "{Supernova emission of secretly interacting neutrino fluid: Theoretical foundations}",
    eprint = "2307.15122",
    archivePrefix = "arXiv",
    primaryClass = "hep-ph",
    doi = "10.1103/PhysRevD.109.023017",
    journal = "Phys. Rev. D",
    volume = "109",
    number = "2",
    pages = "023017",
    year = "2024"
}

@article{Fiorillo:2023ytr,
    author = "Fiorillo, Damiano F. G. and Raffelt, Georg G. and Vitagliano, Edoardo",
    title = "{Large Neutrino Secret Interactions Have a Small Impact on Supernovae}",
    eprint = "2307.15115",
    archivePrefix = "arXiv",
    primaryClass = "hep-ph",
    doi = "10.1103/PhysRevLett.132.021002",
    journal = "Phys. Rev. Lett.",
    volume = "132",
    number = "2",
    pages = "021002",
    year = "2024"
}

@article{Diamond:2023scc,
    author = "Diamond, Melissa and Fiorillo, Damiano F. G. and Marques-Tavares, Gustavo and Vitagliano, Edoardo",
    title = "{Axion-sourced fireballs from supernovae}",
    eprint = "2303.11395",
    archivePrefix = "arXiv",
    primaryClass = "hep-ph",
    doi = "10.1103/PhysRevD.107.103029",
    journal = "Phys. Rev. D",
    volume = "107",
    number = "10",
    pages = "103029",
    year = "2023",
    note = "[Erratum: Phys.Rev.D 108, 049902 (2023)]"
}

@article{Carenza:2023old,
    author = "Carenza, Pierluca and Lucente, Giuseppe and Mastrototaro, Leonardo and Mirizzi, Alessandro and Serpico, Pasquale Dario",
    title = "{Comprehensive constraints on heavy sterile neutrinos from core-collapse supernovae}",
    eprint = "2311.00033",
    archivePrefix = "arXiv",
    primaryClass = "hep-ph",
    reportNumber = "LAPTH-054/23, CA21106",
    doi = "10.1103/PhysRevD.109.063010",
    journal = "Phys. Rev. D",
    volume = "109",
    number = "6",
    pages = "063010",
    year = "2024"
}

@article{Akita:2023iwq,
    author = "Akita, Kensuke and Im, Sang Hui and Masud, Mehedi and Yun, Seokhoon",
    title = "{Limits on heavy neutral leptons, Z' bosons and majorons from high-energy supernova neutrinos}",
    eprint = "2312.13627",
    archivePrefix = "arXiv",
    primaryClass = "hep-ph",
    reportNumber = "CTPU-PTC-23-55",
    doi = "10.1007/JHEP07(2024)057",
    journal = "JHEP",
    volume = "07",
    pages = "057",
    year = "2024"
}

@article{Lella:2024dmx,
    author = "Lella, Alessandro and Ravensburg, Eike and Carenza, Pierluca and Marsh, M. C. David",
    title = "{Supernova limits on QCD axionlike particles}",
    eprint = "2405.00153",
    archivePrefix = "arXiv",
    primaryClass = "hep-ph",
    doi = "10.1103/PhysRevD.110.043019",
    journal = "Phys. Rev. D",
    volume = "110",
    number = "4",
    pages = "043019",
    year = "2024"
}

@article{Fiorillo:2024upk,
    author = "Fiorillo, Damiano F. G. and Vitagliano, Edoardo",
    title = "{Self-Interacting Dark Sectors in Supernovae Can Behave as a Relativistic Fluid}",
    eprint = "2404.07714",
    archivePrefix = "arXiv",
    primaryClass = "hep-ph",
    doi = "10.1103/PhysRevLett.133.251004",
    journal = "Phys. Rev. Lett.",
    volume = "133",
    number = "25",
    pages = "251004",
    year = "2024"
}

@article{Telalovic:2024cot,
    author = "Telalovic, Bernanda and Fiorillo, Damiano F. G. and Mart{\'\i}nez-Mirav{\'e}, Pablo and Vitagliano, Edoardo and Bustamante, Mauricio",
    title = "{The next galactic supernova can uncover mass and couplings of particles decaying to neutrinos}",
    eprint = "2406.15506",
    archivePrefix = "arXiv",
    primaryClass = "hep-ph",
    doi = "10.1088/1475-7516/2024/11/011",
    journal = "JCAP",
    volume = "11",
    pages = "011",
    year = "2024"
}

@article{Benabou:2024jlj,
    author = "Benabou, Joshua N. and Manzari, Claudio Andrea and Park, Yujin and Prabhakar, Garima and Safdi, Benjamin R. and Savoray, Inbar",
    title = "{Time-delayed gamma-ray signatures of heavy axions from core-collapse supernovae}",
    eprint = "2412.13247",
    archivePrefix = "arXiv",
    primaryClass = "hep-ph",
    doi = "10.1103/PhysRevD.111.095029",
    journal = "Phys. Rev. D",
    volume = "111",
    number = "9",
    pages = "095029",
    year = "2025"
}

@article{Alda:2024cxn,
    author = "Alda, Jorge and Levati, Gabriele and Paradisi, Paride and Rigolin, Stefano and Selimovic, Nudzeim",
    title = "{Collider and astrophysical signatures of light scalars with enhanced {\ensuremath{\tau}} couplings}",
    eprint = "2407.18296",
    archivePrefix = "arXiv",
    primaryClass = "hep-ph",
    doi = "10.1007/JHEP06(2025)008",
    journal = "JHEP",
    volume = "06",
    pages = "008",
    year = "2025"
}

@article{Fiorillo:2025yzf,
    author = "Fiorillo, Damiano F. G. and Pitik, Tetyana and Vitagliano, Edoardo",
    title = "{Energy Transfer by Feebly Interacting Particles in Supernovae: The Trapping Regime}",
    eprint = "2503.13653",
    archivePrefix = "arXiv",
    primaryClass = "hep-ph",
    doi = "10.1103/cz94-dqxt",
    journal = "Phys. Rev. Lett.",
    volume = "135",
    number = "7",
    pages = "071005",
    year = "2025"
}

@article{Dev:2023hax,
    author = "Dev, P. S. Bhupal and Fortin, Jean-Fran{\c{c}}ois and Harris, Steven P. and Sinha, Kuver and Zhang, Yongchao",
    title = "{First Constraints on the Photon Coupling of Axionlike Particles from Multimessenger Studies of the Neutron Star Merger GW170817}",
    eprint = "2305.01002",
    archivePrefix = "arXiv",
    primaryClass = "hep-ph",
    reportNumber = "INT-PUB-23-014",
    doi = "10.1103/PhysRevLett.132.101003",
    journal = "Phys. Rev. Lett.",
    volume = "132",
    number = "10",
    pages = "101003",
    year = "2024"
}

@article{Mastrototaro:2019vug,
    author = "Mastrototaro, Leonardo and Mirizzi, Alessandro and Serpico, Pasquale Dario and Esmaili, Arman",
    title = "{Heavy sterile neutrino emission in core-collapse supernovae: Constraints and signatures}",
    eprint = "1910.10249",
    archivePrefix = "arXiv",
    primaryClass = "hep-ph",
    doi = "10.1088/1475-7516/2020/01/010",
    journal = "JCAP",
    volume = "01",
    pages = "010",
    year = "2020"
}

@article{Falk:1978kf,
    author = "Falk, Sydney W. and Schramm, David N.",
    title = "{Limits From Supernovae on Neutrino Radiative Lifetimes}",
    reportNumber = "EFI-78-35-CHICAGO",
    doi = "10.1016/0370-2693(78)90417-3",
    journal = "Phys. Lett. B",
    volume = "79",
    pages = "511",
    year = "1978"
}

@article{Fiorillo:2025sln,
    author = "Fiorillo, Damiano F. G. and Pitik, Tetyana and Vitagliano, Edoardo",
    title = "{Supernova production of axion-like particles coupling to electrons, reloaded}",
    eprint = "2503.15630",
    archivePrefix = "arXiv",
    primaryClass = "hep-ph",
    month = "3",
    year = "2025"
}

@article{Diamond:2023cto,
  title = {Multimessenger Constraints on Radiatively Decaying Axions from GW170817},
  author = {Diamond, M. and Fiorillo, D. and Marques-Tavares, G. and Tamborra, I. and Vitagliano, E.},
  journal = {Phys. Rev. Lett.},
  volume = {132},
  issue = {10},
  pages = {101004},
  numpages = {8},
  year = {2024},
  month = {Mar},
  publisher = {American Physical Society},
  doi = {10.1103/PhysRevLett.132.101004},
  url = {https://link.aps.org/doi/10.1103/PhysRevLett.132.101004}
}

@article{Diamond:2021ekg,
    author = "Diamond, Melissa D. and Marques-Tavares, Gustavo",
    title = "{\ensuremath{\gamma}-Ray Flashes from Dark Photons in Neutron Star Mergers}",
    eprint = "2106.03879",
    archivePrefix = "arXiv",
    primaryClass = "hep-ph",
    doi = "10.1103/PhysRevLett.128.211101",
    journal = "Phys. Rev. Lett.",
    volume = "128",
    number = "21",
    pages = "211101",
    year = "2022"
}

@article{Bollig:2020xdr,
    author = "Bollig, Robert and DeRocco, William and Graham, Peter W. and Janka, Hans-Thomas",
    title = "{Muons in Supernovae: Implications for the Axion-Muon Coupling}",
    eprint = "2005.07141",
    archivePrefix = "arXiv",
    primaryClass = "hep-ph",
    doi = "10.1103/PhysRevLett.125.051104",
    journal = "Phys. Rev. Lett.",
    volume = "125",
    number = "5",
    pages = "051104",
    year = "2020",
    note = "[Erratum: \href{https://doi.org/10.1103/PhysRevLett.126.189901}{\textit{Phys.Rev.Lett.} \textbf{126}, 189901 (2021)}]"
}

@article{Wilks:1938dza,
    author = "Wilks, S. S.",
    title = "{The Large-Sample Distribution of the Likelihood Ratio for Testing Composite Hypotheses}",
    doi = "10.1214/aoms/1177732360",
    journal = "Annals Math. Statist.",
    volume = "9",
    number = "1",
    pages = "60--62",
    year = "1938"
}

@article{Fiorillo:2023frv,
    author = "Fiorillo, Damiano F. G. and Heinlein, Malte and Janka, Hans-Thomas and Raffelt, Georg and Vitagliano, Edoardo and Bollig, Robert",
    title = "{Supernova simulations confront SN 1987A neutrinos}",
    eprint = "2308.01403",
    archivePrefix = "arXiv",
    primaryClass = "astro-ph.HE",
    doi = "10.1103/PhysRevD.108.083040",
    journal = "Phys. Rev. D",
    volume = "108",
    number = "8",
    pages = "083040",
    year = "2023"
}

@ARTICLE{1994ApJ...437L.115I,
       author = {{Iwamoto}, Kohichi and {Nomoto}, Ken'ichi and {H{\"o}flich}, Peter and {Yamaoka}, Hitoshi and {Kumagai}, Shiomi and {Shigeyama}, Toshikazu},
        title = "{Theoretical Light Curves for the Type IC Supernova SN 1994I}",
      journal = {Astrophys. J. Lett.},
         year = 1994,
        month = dec,
       volume = {437},
        pages = {L115},
          doi = {10.1086/187696},
       adsurl = {https://ui.adsabs.harvard.edu/abs/1994ApJ...437L.115I}
}

@article{Yoon:2010at,
    author = "Yoon, Sung-Chul and Woosley, Stan E. and Langer, Norbert",
    title = "{Type Ib/c supernovae in binary systems I. Evolution and properties of the progenitor stars}",
    eprint = "1004.0843",
    archivePrefix = "arXiv",
    primaryClass = "astro-ph.SR",
    doi = "10.1088/0004-637X/725/1/940",
    journal = "Astrophys. J.",
    volume = "725",
    pages = "940--954",
    year = "2010"
}

@article{Woosley:1998hk,
    author = "Woosley, S. E. and Eastman, Ronald G. and Schmidt, Brian P.",
    title = "{Gamma-ray bursts and type Ic supernovae: SN 1998bw}",
    eprint = "astro-ph/9806299",
    archivePrefix = "arXiv",
    reportNumber = "WOO-98-01",
    doi = "10.1086/307131",
    journal = "Astrophys. J.",
    volume = "516",
    pages = "788",
    year = "1999"
}

@ARTICLE{2019ApJ...878...49W,
       author = {{Woosley}, S.~E.},
        title = "{The Evolution of Massive Helium Stars, Including Mass Loss}",
      journal = {Astrophys. J.},
         year = 2019,
        month = jun,
       volume = {878},
       number = {1},
          eid = {49},
        pages = {49},
          doi = {10.3847/1538-4357/ab1b41},
archivePrefix = {arXiv},
       eprint = {1901.00215},
 primaryClass = {astro-ph.SR},
       adsurl = {https://ui.adsabs.harvard.edu/abs/2019ApJ...878...49W}
}

@ARTICLE{1998Natur.395..670G,
       author = {{Galama}, T.~J. and {Vreeswijk}, P.~M. and {van Paradijs}, J. and {Kouveliotou}, C. and {Augusteijn}, T. and {B{\"o}hnhardt}, H. and {Brewer}, J.~P. and {Doublier}, V. and {Gonzalez}, J.-F. and {Leibundgut}, B. and {Lidman}, C. and {Hainaut}, O.~R. and {Patat}, F. and {Heise}, J. and {in't Zand}, J. and {Hurley}, K. and {Groot}, P.~J. and {Strom}, R.~G. and {Mazzali}, P.~A. and {Iwamoto}, K. and {Nomoto}, K. and {Umeda}, H. and {Nakamura}, T. and {Young}, T.~R. and {Suzuki}, T. and {Shigeyama}, T. and {Koshut}, T. and {Kippen}, M. and {Robinson}, C. and {de Wildt}, P. and {Wijers}, R.~A.~M.~J. and {Tanvir}, N. and {Greiner}, J. and {Pian}, E. and {Palazzi}, E. and {Frontera}, F. and {Masetti}, N. and {Nicastro}, L. and {Feroci}, M. and {Costa}, E. and {Piro}, L. and {Peterson}, B.~A. and {Tinney}, C. and {Boyle}, B. and {Cannon}, R. and {Stathakis}, R. and {Sadler}, E. and {Begam}, M.~C. and {Ianna}, P.},
        title = "{An unusual supernova in the error box of the {\ensuremath{\gamma}}-ray burst of 25 April 1998}",
      journal = {Nature},
         year = 1998,
        month = oct,
       volume = {395},
       number = {6703},
        pages = {670-672},
          doi = {10.1038/27150},
archivePrefix = {arXiv},
       eprint = {astro-ph/9806175},
 primaryClass = {astro-ph},
       adsurl = {https://ui.adsabs.harvard.edu/abs/1998Natur.395..670G}
}

@article{Soderberg:2004ht,
    author = "Soderberg, A. M. and others",
    title = "{The Sub-energetic GRB 031203 as a cosmic analogue to GRB 980425}",
    eprint = "astro-ph/0408096",
    archivePrefix = "arXiv",
    reportNumber = "SLAC-PUB-10678",
    doi = "10.1038/nature02757",
    journal = "Nature",
    volume = "430",
    pages = "648",
    year = "2004"
}

@article{Mirabal:2022uml,
    author = "Mirabal, Nestor",
    title = "{Revealing ultra-high-energy cosmic ray acceleration with multi-messenger observations of the nearby GRB 980425/SN 1998bw}",
    eprint = "2210.10822",
    archivePrefix = "arXiv",
    primaryClass = "astro-ph.HE",
    doi = "10.1088/1475-7516/2023/02/060",
    journal = "JCAP",
    volume = "02",
    pages = "060",
    year = "2023"
}

@book{zhang2019physics,
  title={The physics of gamma-ray bursts},
  author={Zhang, Bing},
  year={2019},
  publisher={Cambridge University Press}
}

@article{Goodman:1986az,
    author = "Goodman, J.",
    title = "{Are gamma-ray bursts optically thick?}",
    doi = "10.1086/184741",
    journal = "Astrophys. J. Lett.",
    volume = "308",
    pages = "L47--L50",
    year = "1986"
}

@article{Ghosh:2025rjh,
    author = "Ghosh, Oindrila and Jacobsen, Sunniva and Linden, Tim",
    title = "{Heavy Axions Can Disrupt $\gamma$-ray Bursts}",
    eprint = "2501.08978",
    archivePrefix = "arXiv",
    primaryClass = "astro-ph.HE",
    month = "1",
    year = "2025"
}

@article{Manzari:2024jns,
    author = "Manzari, Claudio Andrea and Park, Yujin and Safdi, Benjamin R. and Savoray, Inbar",
    title = "{Supernova Axions Convert to Gamma Rays in Magnetic Fields of Progenitor Stars}",
    eprint = "2405.19393",
    archivePrefix = "arXiv",
    primaryClass = "hep-ph",
    doi = "10.1103/PhysRevLett.133.211002",
    journal = "Phys. Rev. Lett.",
    volume = "133",
    number = "21",
    pages = "211002",
    year = "2024"
}

@article{Fermi-LAT:2009ihh,
    author = "Atwood, W. B. and others",
    collaboration = "Fermi-LAT",
    title = "{The Large Area Telescope on the Fermi Gamma-ray Space Telescope Mission}",
    eprint = "0902.1089",
    archivePrefix = "arXiv",
    primaryClass = "astro-ph.IM",
    reportNumber = "SLAC-PUB-13620",
    doi = "10.1088/0004-637X/697/2/1071",
    journal = "Astrophys. J.",
    volume = "697",
    pages = "1071--1102",
    year = "2009"
}

@article{Meyer:2020vzy,
    author = "Meyer, Manuel and Petrushevska, Tanja",
    title = "{Search for Axionlike-Particle-Induced Prompt $\gamma$-Ray Emission from Extragalactic Core-Collapse Supernovae with the $Fermi$ Large Area Telescope}",
    eprint = "2006.06722",
    archivePrefix = "arXiv",
    primaryClass = "astro-ph.HE",
    doi = "10.1103/PhysRevLett.124.231101",
    journal = "Phys. Rev. Lett.",
    volume = "124",
    number = "23",
    pages = "231101",
    year = "2020",
    note = "[Erratum: Phys.Rev.Lett. 125, 119901 (2020)]"
}

@article{Candon:2024eah,
    author = "Cand{\'o}n, Francisco R. and Fiorillo, Damiano F. G. and Lucente, Giuseppe and Vitagliano, Edoardo and Vogel, Julia K.",
    title = "{NuSTAR Bounds on Radiatively Decaying Particles from M82}",
    eprint = "2412.03660",
    archivePrefix = "arXiv",
    primaryClass = "hep-ph",
    doi = "10.1103/PhysRevLett.134.171004",
    journal = "Phys. Rev. Lett.",
    volume = "134",
    number = "17",
    pages = "171004",
    year = "2025"
}

@article{Chen:2024ekh,
    author = "Chen, Yu-Xuan and Lei, Lei and Xia, Zi-Qing and Wang, Ziwei and Tsai, Yue-Lin Sming and Fan, Yi-Zhong",
    title = "{Searching for Axionlike Particles with X-Ray Observations of Alpha Centauri}",
    eprint = "2410.16065",
    archivePrefix = "arXiv",
    primaryClass = "astro-ph.HE",
    doi = "10.1103/wy1x-1lh7",
    journal = "Phys. Rev. Lett.",
    volume = "134",
    number = "24",
    pages = "241001",
    year = "2025"
}

@article{Muller:2023pip,
    author = {M{\"u}ller, Eike and Carenza, Pierluca and Eckner, Christopher and Goobar, Ariel},
    title = "{Constraining MeV-scale axionlike particles with Fermi-LAT observations of SN 2023ixf}",
    eprint = "2306.16397",
    archivePrefix = "arXiv",
    primaryClass = "astro-ph.HE",
    reportNumber = "LAPTH-038/23",
    doi = "10.1103/PhysRevD.109.023018",
    journal = "Phys. Rev. D",
    volume = "109",
    number = "2",
    pages = "023018",
    year = "2024"
}

@article{Barbarino:2017mes,
    author = "Barbarino, C. and others",
    title = "{LSQ14efd: observations of the cooling of a shock break-out event in a type Ic Supernova}",
    eprint = "1707.04644",
    archivePrefix = "arXiv",
    primaryClass = "astro-ph.HE",
    doi = "10.1093/mnras/stx1709",
    journal = "Mon. Not. Roy. Astron. Soc.",
    volume = "471",
    number = "2",
    pages = "2463--2480",
    year = "2017"
}

@article{YoungSupernovaExperiment:2021fur,
    author = "Gagliano, Alexander and others",
    collaboration = "Young Supernova Experiment",
    title = "{An Early-time Optical and Ultraviolet Excess in the Type-Ic SN 2020oi}",
    eprint = "2105.09963",
    archivePrefix = "arXiv",
    primaryClass = "astro-ph.HE",
    doi = "10.3847/1538-4357/ac35ec",
    journal = "Astrophys. J.",
    volume = "924",
    number = "2",
    pages = "55",
    year = "2022"
}

@article{Muller:2023vjm,
    author = {M{\"u}ller, Eike and Calore, Francesca and Carenza, Pierluca and Eckner, Christopher and Marsh, M. C. David},
    title = "{Investigating the gamma-ray burst from decaying MeV-scale axion-like particles produced in supernova explosions}",
    eprint = "2304.01060",
    archivePrefix = "arXiv",
    primaryClass = "astro-ph.HE",
    doi = "10.1088/1475-7516/2023/07/056",
    journal = "JCAP",
    volume = "07",
    pages = "056",
    year = "2023"
}

@article{Fiorillo:2025gnd,
    author = "Fiorillo, Damiano F. G. and Gil Muyor, {\'A}ngel and Janka, Hans-Thomas and Raffelt, Georg G. and Vitagliano, Edoardo",
    title = "{Axion-photon conversion in transient compact stars: Systematics, constraints, and opportunities}",
    eprint = "2509.13322",
    archivePrefix = "arXiv",
    primaryClass = "hep-ph",
    month = "9",
    year = "2025"
}

@article{Tinyanont:2016vbz,
    author = "Tinyanont, Samaporn and others",
    title = "{A Systematic Study of Mid-Infrared Emission from Core-Collapse Supernovae with SPIRITS}",
    eprint = "1601.03440",
    archivePrefix = "arXiv",
    primaryClass = "astro-ph.SR",
    doi = "10.3847/1538-4357/833/2/231",
    journal = "Astrophys. J.",
    volume = "833",
    number = "2",
    pages = "231",
    year = "2016"
}

@article{Prentice:2018ual,
    author = "Prentice, S. J. and others",
    title = "{Investigating the properties of stripped-envelope supernovae, what are the implications for their progenitors?}",
    eprint = "1812.03716",
    archivePrefix = "arXiv",
    primaryClass = "astro-ph.HE",
    doi = "10.1093/mnras/sty3399",
    journal = "Mon. Not. Roy. Astron. Soc.",
    volume = "485",
    number = "2",
    pages = "1559--1578",
    year = "2019"
}

@article{Kankare:2021fzf,
    author = "Kankare, E. and others",
    title = "{Core-collapse supernova subtypes in luminous infrared galaxies}",
    eprint = "2102.13512",
    archivePrefix = "arXiv",
    primaryClass = "astro-ph.SR",
    doi = "10.1051/0004-6361/202039240",
    journal = "Astron. Astrophys.",
    volume = "649",
    pages = "A134",
    year = "2021"
}

@article{Barbarino:2020amq,
    author = "Barbarino, C. and others",
    title = "{Type Ic supernovae from the (intermediate) Palomar Transient Factory}",
    eprint = "2010.08392",
    archivePrefix = "arXiv",
    primaryClass = "astro-ph.SR",
    doi = "10.1051/0004-6361/202038890",
    journal = "Astron. Astrophys.",
    volume = "651",
    pages = "A81",
    year = "2021"
}

@article{Calore:2021lih,
    author = "Calore, Francesca and Carenza, Pierluca and Giannotti, Maurizio and Jaeckel, Joerg and Lucente, Giuseppe and Mastrototaro, Leonardo and Mirizzi, Alessandro",
    title = "{511~keV line constraints on feebly interacting particles from supernovae}",
    eprint = "2112.08382",
    archivePrefix = "arXiv",
    primaryClass = "hep-ph",
    doi = "10.1103/PhysRevD.105.063026",
    journal = "Phys. Rev. D",
    volume = "105",
    number = "6",
    pages = "063026",
    year = "2022"
}

@article{Sung:2019xie,
    author = "Sung, Allan and Tu, Huitzu and Wu, Meng-Ru",
    title = "{New constraint from supernova explosions on light particles beyond the Standard Model}",
    eprint = "1903.07923",
    archivePrefix = "arXiv",
    primaryClass = "hep-ph",
    doi = "10.1103/PhysRevD.99.121305",
    journal = "Phys. Rev. D",
    volume = "99",
    number = "12",
    pages = "121305",
    year = "2019"
}

@article{Lecce:2025vjc,
    author = "Lecce, Francesca and Lella, Alessandro and Lucente, Giuseppe and Giannotti, Maurizio and Mirizzi, Alessandro",
    title = "{Detecting light axions from supernovae in nearby galaxies}",
    eprint = "2512.04185",
    archivePrefix = "arXiv",
    primaryClass = "hep-ph",
    reportNumber = "BARI-TH/781-25",
    month = "12",
    year = "2025"
}

@article{Ma:2025css,
    author = "Ma, Xiaoran and others",
    title = "{Supernovae at distances {\ensuremath{<}}40 Mpc - I. Catalogues and fractions of supernovae in a complete sample}",
    eprint = "2504.04393",
    archivePrefix = "arXiv",
    primaryClass = "astro-ph.HE",
    doi = "10.1051/0004-6361/202452684",
    journal = "Astron. Astrophys.",
    volume = "698",
    pages = "A305",
    year = "2025"
}

\onecolumngrid
\appendix

\setcounter{equation}{0}
\setcounter{figure}{0}
\setcounter{table}{0}
\setcounter{page}{1}
\makeatletter
\renewcommand{\theequation}{S\arabic{equation}}
\renewcommand{\thefigure}{S\arabic{figure}}
\renewcommand{\thepage}{S\arabic{page}}

\begin{center}
\textbf{Supplemental Material for the Letter\\
\textit{``Small Progenitors, Large Couplings:
Type Ic Supernova Constraints on Radiatively Decaying Particles''}}

\end{center}

\vspace{0.5em}

In this Supplemental Material, we discuss the close analogy between decay bounds from SNe and beam dumps. We also review the production and subsequent decay of ALPs from SNe, including the potential formation of a fireball for large enough coupling. Finally, we provide comprehensive information on the SNe used in this work and on the statistical analysis adopted.

\bigskip

\twocolumngrid

\begin{figure*}
    \centering
    \includegraphics[width=\columnwidth]{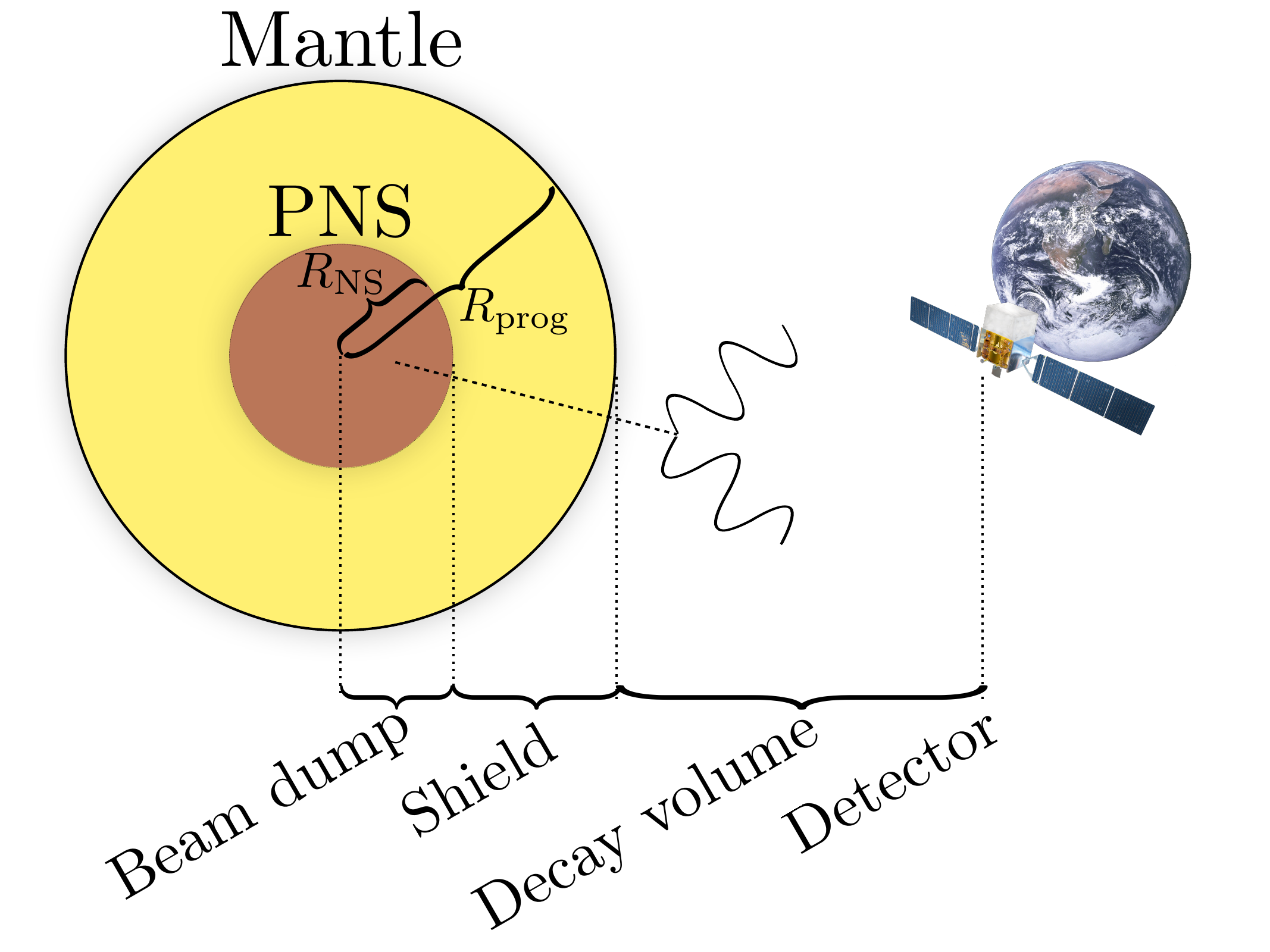}
    \includegraphics[width=\columnwidth]{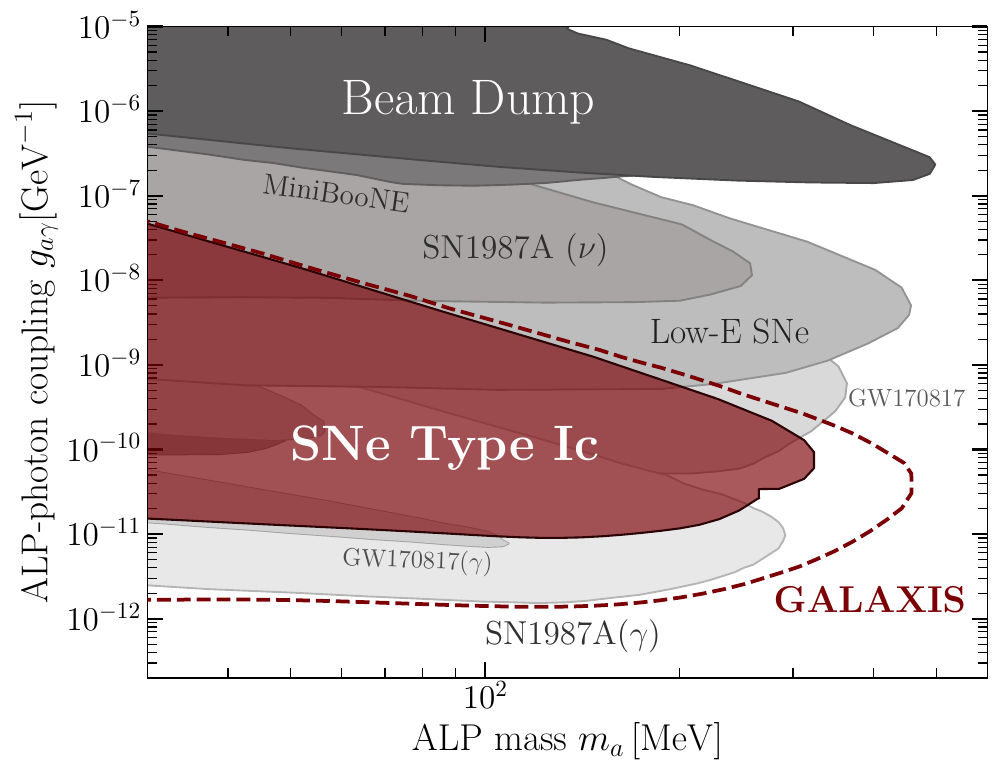}
\caption{Decay bounds from SNe as a cosmic beam dump: sketch of the geometry (left) and resulting constraints together with the GALAXIS projection for a 10 Mpc Type Ic SN using the cold model (right).}\label{fig:cosmic_beam_dump}
\end{figure*}

\section{A.~Type Ic SNe as a cosmic beam dump}
As argued in the main text, it is instructive for a particle physicist to regard the decay bounds from SNe as a cosmic beam dump. Novel particles are produced in the hot, dense core, propagate through the progenitor which acts as a natural shield, impeding the gamma-rays directly produced in the core from reaching Earth, and can subsequently decay outside of the progenitor, leading to a visible gamma-ray flash. This analogy is sketched graphically in Fig.~\ref{fig:cosmic_beam_dump} (left panel). This makes it immediately clear that the best ``beam dumps'' to probe small decay lengths are going to be those with the thinnest possible shield, i.e. those with the smallest $R_{\rm prog}$, directly pointing to Type Ic SNe as an ideal candidate.

This analogy emerges forcefully when comparing the shape of the constraints from Type Ic SNe with that of terrestrial beam dumps, especially in their ceiling, as we do in Fig.~\ref{fig:cosmic_beam_dump} (right panel). For both cases, the constraint closes along a line corresponding to a constant decay length for the particle, roughly equal to a fraction of the length of the shield. For Type Ic SNe, the length of the shield is roughly $L\sim 10^{10}\,\mathrm{cm}$, while the typical energy is of the order of $E_a\sim 100\,\mathrm{MeV}$. For terrestrial beam dumps, the length of the shield is instead of the order of $L\sim 10^4\,\mathrm{cm}$, with a typical energy of $E_a\sim 10\,\mathrm{GeV}$. Since the decay length scales in proportion to $\ell_a \sim 64 \pi E_a/g_{a\gamma}^2$, we find that the constraint on $g_{a\gamma}$ should scale as $g_{a\gamma}\propto \sqrt{E_a/L}$; therefore, by this naive estimate, we expect constraints from Type Ic SNe to show their ceiling at a coupling about four orders of magnitude lower than for terrestrial beam dumps. Such a naive estimate is in surprising good agreement with the constraints in the right panel of Fig.~\ref{fig:cosmic_beam_dump}.

Notice that in Fig.~\ref{fig:cosmic_beam_dump}, we do not show the recently proposed constraints from short gamma-ray bursts (GRBs)~\cite{Ghosh:2025rjh}. These were obtained assuming an internal radiation-dominated fireball with typical temperatures $T\sim 100\,\mathrm{MeV}$, a number inspired, under extremely optimistic conditions, by the original Goodman model~\cite{Goodman:1986az} in which all the GRB energy $\mathcal{E}\sim 10^{54}\,\mathrm{erg}$ is instantaneously injected within an engine radius $r_s\sim 10^6\,\mathrm{cm}$, leading to a temperature $T\sim(3\mathcal{E}/4\pi r_s^3)^{1/4}\sim 100\,\mathrm{MeV}$. However, the duration of a GRB is in reality $\delta t\sim 1\,\mathrm{s}\gg r_s$, so instantaneous injection is not a realistic picture. All modern treatments of GRBs (see e.g. the textbook treatment of Ref.~\cite{zhang2019physics}) assume therefore a steady injection of energy over a time $\delta t$, implying a central temperature $T\sim (\mathcal{E}/4\pi r_s^2\delta t)^{1/4}\sim 1 - 10\,\mathrm{MeV}$. The temperature is likely even lower, since at the base of the jet launched by the engine most of the energy is likely in the form of Poynting flux rather than thermal.  With a temperature of $1\,\mathrm{MeV}$, lower than their assumed $100\,\mathrm{MeV}$ by two orders of magnitude, the excluded area in their Fig.~4 would shrink down to masses two orders of magnitude lower due to kinematical suppression by the Boltzmann factor $e^{-m_a/T}$. Even in the low-mass regime, the excluded region would be completely disrupted; since the energy emission rate by Primakoff grows as $\propto T^4$, as we also review below, there is a corresponding change in the cooling rate by $\sim 8$ orders of magnitude, resulting in $\sim 4$ orders of magnitude in the excluded coupling. Hence, since the temperature is so much lower than assumed in Ref.~\cite{Ghosh:2025rjh}, the GRB constraints disappear completely.

In addition to the constraints derived in the main text, we present a projection for the prospective detection of an extragalactic Type Ic SN in the most optimistic scenario of a $4\pi$ instantaneous field of view gamma-ray instrument. Assuming an effective area comparable to that of Fermi-LAT in our energy range of interest (around $100 \, \mathrm{MeV}$), we estimate the projected sensitivity by rescaling one representative event from our sample; we choose SN2020oi. For this estimate, we adopt a fiducial distance of 10 Mpc, which is optimistic but plausible (the closest SN in our sample lies at 11.6 Mpc). An instrument such as the proposed GALAXIS \cite{Manzari:2024jns}, based on a constellation of SmallSats, is expected to provide near-continuous coverage: multiple spacecraft would monitor the same sky region simultaneously, so that temporary interruptions affecting individual satellites (e.g., due to South Atlantic Anomaly passages or safing events) would not result in gaps for the source. Accordingly, we assume continuous sensitivity and consider only the single one-hour time bin within the $v_1$ that maximizes the exposure (effective area times effective observing time), which we take to be approximately constant under the considerations above. In the notation of our main analysis, this corresponds to $P_{i,\beta} = 1$, where $\beta$ denotes the single maximum exposure bin and $i$ denotes SN2020oi, reducing to the standard Poisson likelihood case. For that one-hour bin we take $N = 125$ observed counts and  $B = 160$ expected background counts. The resulting projection is shown in Fig.~\ref{fig:cosmic_beam_dump} as a red dashed line.

\section{B.~ALP production in SNe cores}

The production of ALPs coupling to photons within the cores of SNe happens through Primakoff conversion $\gamma+N\to a+N$ and coalescence $\gamma+\gamma\to a$. We have recently reviewed the features of these production processes in Refs.~\cite{Diamond:2023cto,Fiorillo:2025yzf}, to which we refer for a more detailed discussion. Here we limit ourselves to providing the resulting volumetric axion emissivity; notation-wise, we will denote number densities with small Latin letters $n_a$, volume-integrated numbers with capital Latin letters $N_a$, and emission rates with a dot, so the volumetric emission rate is $\dot{n}_a$ and the total emission rate is $\dot{N}_a$. The differential emission rates in the axion energy $E_a$ are explicitly denoted as $d\dot{n}_a/dE_a$ and $d\dot{N}_a/dE_a$.

For both processes contributing to the emission, we can write the emission rate as
\begin{equation}
    \frac{d\dot{n}_a}{dE_a}=\frac{E_a p_a}{2\pi^2 (e^{E_a/T}-1)}\tilde{\gamma}_a,
\end{equation}
where $p_a=\sqrt{E_a^2-m_a^2}$ is the axion momentum, expressed in terms of its mass $m_a$, and $\tilde{\gamma}_a$ is the reduced absorption coefficient, encoding the information on the interaction. 

For Primakoff emission, the reduced absorption coefficient reads
\begin{equation}\label{eq:gammatilde_primakoff}
    \tilde{\gamma}_{a}=2\hat{n}\frac{E_a}{p_a}\frac{Z^2\alpha g_{a\gamma}^2}{2}f_P,
\end{equation}
where $Z$ is the atomic number (we take $Z=1$), $\hat{n}=Y_e n_B$ is the charge density (in our case $\hat{n}=n_e=n_p$ is equal to the electron and proton number density), and $g_{a\gamma}$ is the axion photon coupling. Compared with Ref.~\cite{Fiorillo:2025yzf}, we do \textit{not} include small corrections due to the plasma frequency of the photons; as recently argued in Ref.~\cite{Fiorillo:2025gnd}, these corrections are anyway not consistently implemented without considering the contribution of longitudinal modes and other effects of the same order of magnitude. Since these are all negligible, we simply do not consider them. Finally, we have
    \begin{widetext}
\begin{equation}
    f_P=\frac{[(E_a+p_a)^2+k_S^2][(E_a-p_a)^2+k_S^2]}{16k_S^2 E_ap_a}\log\frac{(E_a+p_a)^2+k_S^2}{(E_a-p_a)^2+k_S^2}-\frac{(E_a^2-p_a^2)^2}{16k_S^2 E_a p_a}\log\frac{(E_a+p_a)^2}{(E_a-p_a)^2}-\frac{1}{4},
\end{equation}
\end{widetext}
where the Debye screening scale is $k_S^2=4\pi\alpha\hat{n}/T$ and $T$ is the temperature. In the absence of heavy elements, this gives
\begin{equation}
    \tilde{\gamma}_a=5.46\times 10^{18}\; \mathrm{s}^{-1}\;\frac{Y_e \rho}{10^{14}\;\mathrm{g/cm}^3}\;\left(\frac{g_{a\gamma}}{1\;\mathrm{GeV}^{-1}}\right)^2\frac{k}{p}f_P.
\end{equation}
As in Ref.~\cite{Caputo:2022mah}, we extract $Y_e$ assuming that $Y_e=1-X_n$, where $X_n$ is the neutron fraction. 

For coalescence, the reduced absorption rate comes from the inverse process, i.e. ALP decay. We then have
\begin{equation}
    \tilde{\gamma}_a=\frac{g_{a\gamma}^2 m_a^4 \tilde{f}_B}{64 \pi E_a},
\end{equation}
with
\begin{equation}
    \tilde{f}_B=\frac{2T}{p_a}\log\left[\frac{e^{\frac{E_a+p_a}{4T}}-e^{-\frac{E_a+p_a}{4T}}}{e^{\frac{E_a-p_a}{4T}}-e^{-\frac{E_a-p_a}{4T}}}\right].
\end{equation}

If $T\ll E_a, m_a$, this simplifies to $f_B\to 1$, i.e. decay in vacuum. We can also write
\begin{equation}
    \tilde{\gamma}_a=7.46\times 10^{18}\;\mathrm{s}^{-1}\;\left(\frac{g_{a\gamma}}{1\;\mathrm{GeV}^{-1}}\right)^2\;\left(\frac{m_a}{100\;\mathrm{MeV}}\right)^3 \frac{m_a}{E_a}f_B.
\end{equation}
These expressions completely characterize the volumetric emissivity, which can then be integrated over the volume of the SN profile to obtain the final ALP flux. To account for the gravitational lapse, we evaluate the differential rate $d\dot{n}_a/dE_a$ at the local energy $E_{a,\rm loc}=E_a (1+z)$. In this way, the total number of particles emitted per unit time transforms as $\dot{n}_a=\int dE_a d\dot{n}_a/dE_a[E_{a,\rm loc}] = \dot{n}_{a,\rm loc}/(1+z)$, as it pertains to a rate due to the time dilation. Similarly, the integrated luminosity $L_a=\int E_a dE_a d\dot{n}_a/dE_a$ scales with $(1+z)^{-2}$. The ALP momentum is $p_{a,\rm loc}=\sqrt{E_{a,\rm loc}^2-m_a^2}$; particles with energies in the observer frame $E_a<m_a$ cannot of course escape to infinity and are gravitationally trapped. By integrating over the volume and duration of the SN simulation, we finally obtain the total number of emitted ALPs $dN_a/dE_a$.

\begin{figure}
    \centering
    \includegraphics[width=\columnwidth]{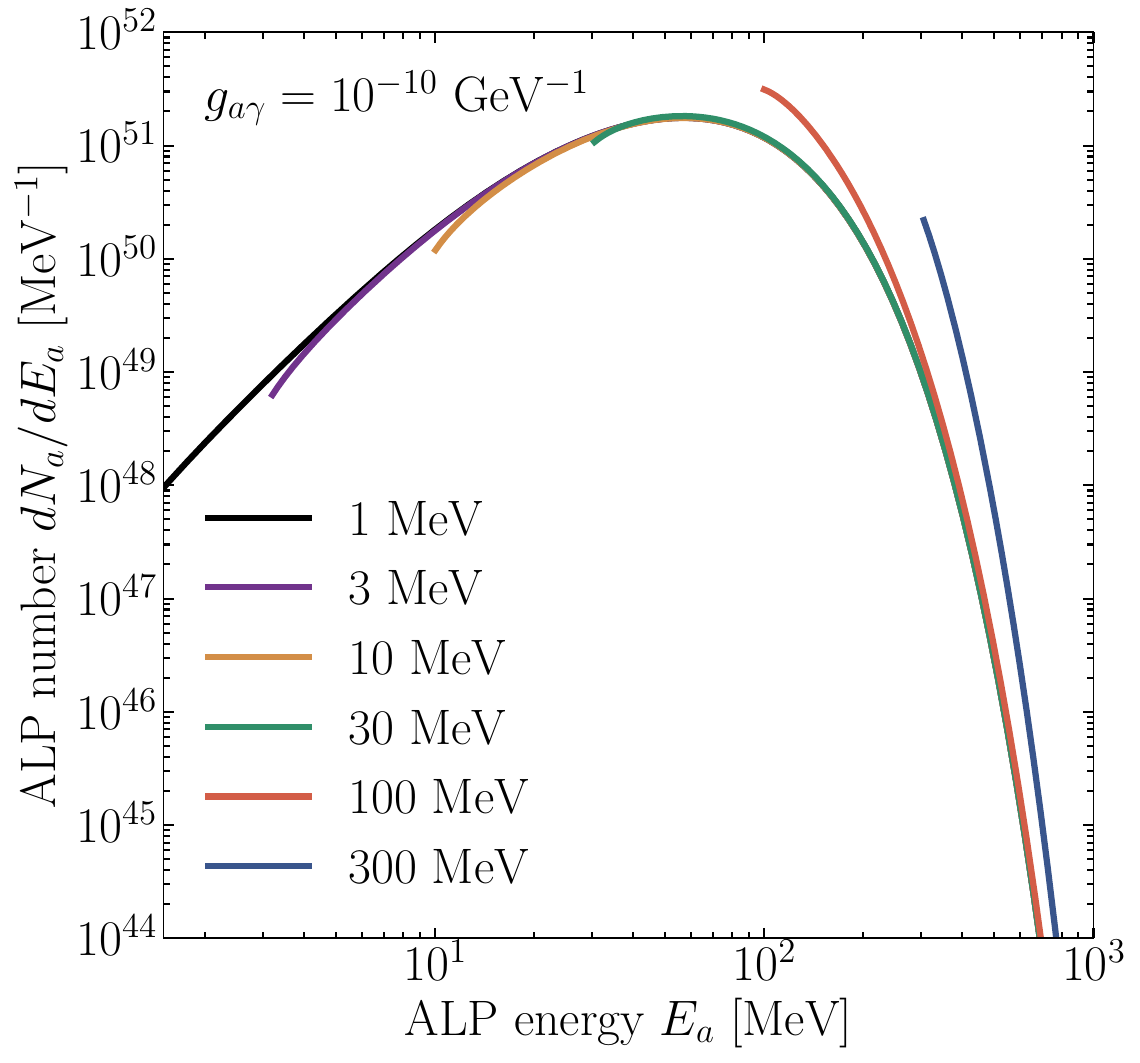}
\caption{Number of emitted ALPs per unit energy, for varying ALP mass (listed in the legend), using the cold SN model. We fix the coupling $g_{a\gamma}=10^{-10}\,\mathrm{GeV}^{-1}$, with the overall number of events scaling as $g_{a\gamma}^2$ in the free-streaming regime.}\label{fig:spectrum}
\end{figure}

For reproducibility, we show in Fig.~\ref{fig:spectrum} the number of emitted ALPs obtained from integration over the cold SN model for varying ALP masses. For low masses ($m_a\sim 1-10\,\mathrm{MeV}$), the emissivity is dominated by Primakoff, and increasing the mass simply changes the minimum energy of the spectrum. For large masses ($m_a\gtrsim 100\,\mathrm{MeV}$), coalescence becomes the dominant emission channel, so that the spectrum, besides moving to larger masses, also increases significantly in normalization.

\section{C.~Fireball formation}

If the density of the photons produced in ALP decay outside of the progenitor is large enough, they will produce pairs via mutual scattering $\gamma+\gamma\to e^++e^-$. These pairs will subsequently radiate photons in bremsstrahlung $e+e\to e+e+\gamma$, which will degrade the energy of the particles, originally at around 100~MeV, into the 100-keV range, forming a fireball~\cite{Diamond:2023scc,Diamond:2023cto}---a shell of hydrodynamic radiation coupled by mutual interactions. The region of ALP coupling and mass leading to fireball formation, which we refer to for brevity as fireball region, produces a completely different signature, peaking in the energy range of X-ray experiments rather than Fermi-LAT. Since this region lies largely in portions of the parameter space that are already excluded by other observables, we do not investigate whether these X-ray experiments can exclude it---although this is still a conceptually relevant task that we plan to pursue in a forthcoming publication. Nonetheless, in this section we identify where this region lies. We follow Ref.~\cite{Diamond:2023scc}, whose results we review here.

The first condition for fireball formation is that the photons produced by ALP decay undergo pair production rapidly enough, i.e. before they advect away. This basically means that the timescale for pair annihilation among the photons should be shorter than the light-crossing time for photons at their typical decay radius. Thus, we need to first determine the total amount of photons produced in ALP decay outside of the progenitor star, and the typical radius at which they are injected. The latter is the average of the decay radius $r_{\rm dec}$ over the energy distribution of the ALPs, integrated however only over the region $r_{\rm dec}>R_{\rm prog}$. If we call 
\begin{equation}
    \ell_a=\frac{p_a}{\Gamma_a m_a}
\end{equation}
the decay length of an ALP with energy $E_a$, the average decay radius outside of the progenitor, which we identify as the fireball radius, is
\begin{equation}
    r_{f}=R_{\rm prog}+\langle \ell_a\rangle,
\end{equation}
where we introduce the average over the ALP distribution of a generic quantity $x$ as

\begin{equation}
    \langle x \rangle= \frac{\int dE_a x e^{-R_{\rm prog}/\ell_a}\frac{dN_a}{dE_a}}{\int dE_a e^{-R_{\rm prog}/\ell_a}\frac{dN_a}{dE_a}}.
\end{equation}

The fireball is characterized by the total amount of ALPs $\mathcal{N}_a$, ALP energy $\mathcal{E}_a$, and ALP momentum $\mathcal{P}_a$ that is released outside of the progenitor; these are all obtained by integrating the distribution of ALPs decaying outside of the progenitor $ e^{-R_{\rm prog}/\ell_a} dN_a/dE_a$ over $dE_a$, $E_a dE_a$, and $p_a dE_a$ respectively. 

The photons are produced by ALP decay into a shell whose width is determined by the spread in the location of decay; this spread depends in turn on the decay probability, on the velocity of the decaying axions, and on the initial width of the axion signal (which we take coarsely of the order of $\Delta_1=3\,\mathrm{s}$). The final result, systematically determined in Ref.~\cite{Diamond:2023scc}, for the average width $\Delta$, is
\begin{widetext}
\begin{equation}
     \Delta^2=\frac{\Delta_1^2 \langle v_a^2\rangle}{12}+\left\langle(1-v_a)^2\left(\frac{(R_{\rm prog}+\ell_a)^2+\ell_a^2}{v_a^2}\right)\right\rangle -\left\langle\frac{(1-v_a)(\ell_a+R_{\rm prog})}{v_a}\right\rangle^2.
\end{equation}
\end{widetext}
The first term measures the spread induced by the initial duration of the ALP signal, while the second term describes the spread induced by the decay.

The number density of photons injected by ALP decay is
\begin{equation}
    n_\gamma=\frac{2\mathcal{N}_a}{4\pi r_f^2 \Delta},
\end{equation}
where the factor $2$ comes from the two photons produced per ALP decay, so that the condition of efficient pair production reads
\begin{equation}\label{eq:efficient_pair}
    n_\gamma \sigma_{\gamma\gamma\to e^+ e^-}\Delta=\frac{2\mathcal{N}_a \sigma_{\gamma\gamma\to e^+ e^-}}{4\pi r_f^2}\gg 1.
\end{equation}
Here $\sigma_{\gamma\gamma\to e^+ e^-}$ is the pair production cross section, which must be evaluated at the typical center-of-mass scattering for two photons in the comoving frame of the photon plasma (i.e. in the frame where $\mathcal{P}'_a$, the total ALP momentum, vanishes). Denoting by primed the variables in this frame, we have that the typical photon energy is $E'_\gamma=m_a/2$; the well-known expression for the Breit-Wheeler pair production cross section is reported, e.g., in Eq. (12) of Ref.~\cite{Diamond:2023scc}. 

If pair production is rapid enough, the plasma will thermalize into an equilibrium state with both pairs and photons, determined by the conditions of thermal and chemical equilibrium. Conservation of energy, particle number, and momentum lead to the inferred temperature, chemical potential (which is the same for photons, electron, and positrons, assuming them to all be ultra-relativistic), and velocity of the fluid, as discussed in Ref.~\cite{Diamond:2023scc}. The bulk velocity of the fluid is
\begin{equation}
    v=\frac{2\mathcal{E}_a}{\mathcal{P}_a}-\sqrt{\frac{4\mathcal{E}_a^2}{\mathcal{P}_a^2}-3},
\end{equation}
with Lorentz factor $\gamma=1/\sqrt{1-v^2}$, and the comoving fluid temperature is
\begin{equation}
    T'_i=\frac{\mathcal{P}_a}{8\gamma v \mathcal{N}_a}.
\end{equation}

Finally, the particle number density in the laboratory frame is
\begin{equation}
    n=\gamma n'=\frac{2\mathcal{N}_a}{4\pi r_f^2 \Delta}.
\end{equation}
The thermalization induced by pair production cannot lead to a change in the number of particles; the latter is instead induced by bremsstrahlung. The condition for efficient bremsstrahlung is
\begin{equation}\label{eq:efficient_bremsstrahlung}
    \frac{2}{3}\frac{2\mathcal{N}_a}{4\pi r_f^2 \Delta}\sigma_{ee\to ee\gamma}(T'_i)\Delta \gg 1,
\end{equation}
where the factor $2/3$ accounts for the fraction of particles of the plasma that are electrically charged, and $\sigma_{ee\to ee\gamma}(T_i)$ is the cross section for bremsstrahlung (taken from Eq. (16) of Ref.~\cite{Diamond:2021ekg}) evaluated at the typical comoving particle energy $T'_i$.

\begin{figure}
    \centering
    \includegraphics[width=\columnwidth]{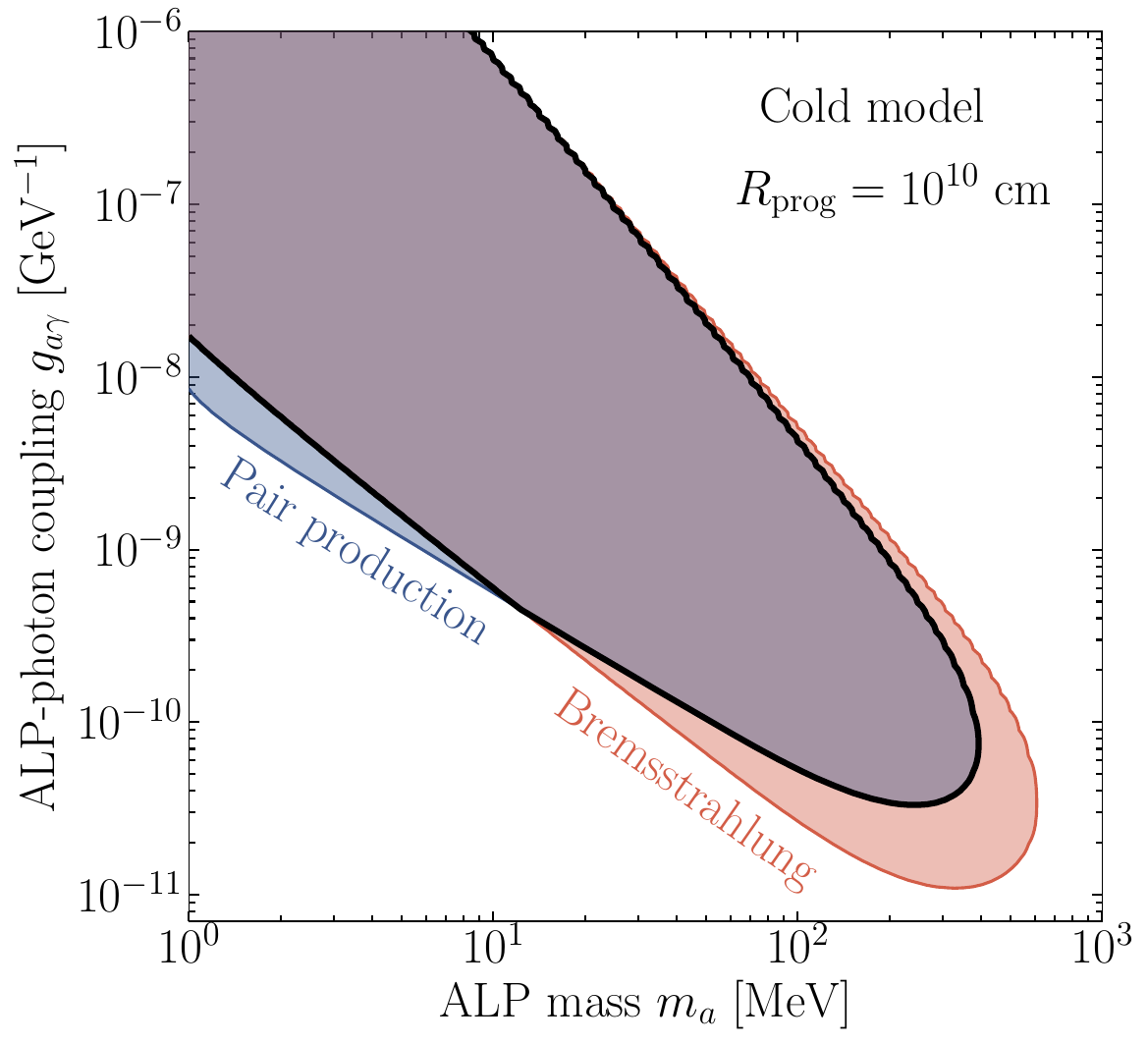}
    \caption{Regions in the parameter space where pair production and bremsstrahlung are rapid enough to thermalize,  using the cold SN model. In the intersection of the two regions (marked by a black line), a fireball is formed with consequent reprocessing of the gamma-ray energy in the 100-keV range.}
    \label{fig:fireball}
\end{figure}

Figure~\ref{fig:fireball} shows the regions of parameter space where the conditions for an efficient pair production (Eq.~\eqref{eq:efficient_pair}) and bremsstrahlung (Eq.~\eqref{eq:efficient_bremsstrahlung}) are satisfied. The intersection of these regions is the fireball region, where the pairs produced by pair production can thermalize and reprocess the overall photon energy in the 100-keV range.

\section{D.~Set of Type Ic SNe used for the analysis}

\begin{table*}
\centering
\footnotesize
\setlength{\tabcolsep}{8pt}
\begin{tabular}{
c |
  c
  c
  c
  c
  c
  c
  | c
}
\hline\hline
\multicolumn{1}{c|}{SN} &
\multicolumn{1}{c}{R.A. [deg]} &
\multicolumn{1}{c}{Dec. [deg]} &
\multicolumn{1}{c}{$t_{\rm coll}$ [MJD]} &
\multicolumn{1}{c}{$t_{\rm coll}$ [UTC]} &
\multicolumn{1}{c}{$D$ [Mpc]} &
\multicolumn{1}{c}{Ref.} &
\multicolumn{1}{c}{$f_i$} \\
\hline
PTF10zcn   & 349.810  &  26.053  & $55464.630^{+2.00}_{-2.00}$ & \texttt{2010-09-25 15:07:12}  & 90.3   & \cite{Barbarino:2020amq} & 0.31 \\
SN2011bm   & 194.225  &  22.375  & $55639.700^{+0.78}_{-0.92}$ & \texttt{2011-03-19 16:48:00}  & 94.2   & \cite{Meyer:2020vzy} & 0.38 \\
PTF12gzk   & 333.173  &   0.512  & $56131.490^{+0.13}_{-0.13}$ & \texttt{2012-07-23 11:45:36}  & 58.7   & \cite{Meyer:2020vzy} & 0.47 \\
PTF13aot   & 199.609  &  31.469  & $56395.550^{+2.00}_{-2.00}$ & \texttt{2013-04-13 13:12:00}  & 80.7   & \cite{Barbarino:2020amq} & 0.31 \\
SN2013dk   & 180.471  & -18.872  & $56492.500^{+5.0}_{-5.0}$   & \texttt{2013-07-19 12:00:00}  & 22.3   & \cite{Tinyanont:2016vbz} & 0.37 \\
PTF13djf   & 353.411  &   8.812  & $56537.860^{+2.00}_{-2.00}$ & \texttt{2013-09-02 20:38:24}  & 87.1   & \cite{Barbarino:2020amq} & 0.35 \\
SN2013ge   & 158.702  &  21.662  & $56595.540^{+1.29}_{-0.66}$ & \texttt{2013-10-30 12:57:36}  & 18.7   & \cite{Meyer:2020vzy} & 0.43 \\
SN2014L    & 184.703  &  14.412  & $56681.330^{+0.31}_{-0.31}$ & \texttt{2014-01-24 07:55:12}  & 34.4   & \cite{Meyer:2020vzy} &  0.43 \\
LSQ14efd   &  53.912  & -58.877  & $56880.500^{+2.00}_{-2.00}$ & \texttt{2014-08-11 12:00:00}  & 287.8  & \cite{Barbarino:2017mes} &  0.30 \\
PTF14gqr   & 353.366  &  33.646  & $56949.880^{+2.00}_{-2.00}$ & \texttt{2014-10-19 21:07:12}  & 77.6   & \cite{Barbarino:2020amq} & 0.35 \\
PTF15dtg   &  37.584  &  37.235  & $57322.520^{+2.29}_{-3.96}$ & \texttt{2015-10-27 12:28:48}  & 224.4  & \cite{Barbarino:2020amq} & 0.21 \\
SN2016iae  &  43.333  & -32.467  & $57694.000^{+4.0}_{-4.0}$   & \texttt{2016-11-02 00:00:00}  & 17.1   & \cite{Prentice:2018ual} & 0.35 \\
SN2017ein  & 178.222  &  44.124  & $57896.360^{+0.60}_{-0.60}$ & \texttt{2017-05-23 08:38:24}  & 11.6   & \cite{Meyer:2020vzy} & 0.41 \\
SN2018ec   & 175.000  &  15.000  & $58090.400^{+5.0}_{-5.0}$   & \texttt{2017-12-03 09:36:00}  & 40.1   & \cite{Kankare:2021fzf} & 0.21 \\
SN2020oi   & 185.729  &  15.824  & $58854.000^{+0.30}_{-0.30}$ & \texttt{2020-01-06 00:00:00}  & 15.5   & \cite{YoungSupernovaExperiment:2021fur} & 0.45 \\

\hline\hline
\end{tabular}
\caption{Sample of Type Ic SNe used in this work. The columns list the equatorial coordinates, the explosion epoch $t_{\rm coll}$ (MJD) with uncertainties, and the corresponding UTC conversion. Distances in Mpc are computed from the host redshift $z$ via the low-$z$ Hubble–Lemaître law $D = cz/H_{0}$, using $c = 299{,}792.458\,\mathrm{km\,s^{-1}}$ and $H_{0} = 70\,\mathrm{km\,s^{-1}\,Mpc^{-1}}$.  Finally, the probability that the gamma-ray flash happens within a time frame in which the experiment was observing $f_i=\sum_{\overline{\alpha}} P_{i,\overline{\alpha}}$, where $\overline{\alpha}$ denotes the bins in which the experiment is actually sensitive, is also shown.}
\label{tab:SNe_sample}
\end{table*}

In Table~\ref{tab:SNe_sample}, we collect the sample of SNe that we use for our statistical analysis. Our primary motivation in selecting this group of events is related to the relatively low uncertainty on the explosion time. Notice that the distance of the SNe can vary quite widely among different events, ranging from 10 Mpc to 300 Mpc roughly. This is not crucial for our constraints, since they primarily improve the results in the trapping regime, where the couplings are so large that the photon flux, if it is able to escape the progenitor star, would be large enough to overshoot the background of the experiment even at very large distances. 

The probability that the experiment is active at the moment of the explosion $f_i$ for the different SNe is roughly around $0.3$, as expected since Fermi-LAT is typically sensitive to about one fifth of the sky. For different sources, this can range from $0.21$ to $0.47$. These typical values motivate our estimate in the main text that $\mathcal{O}(10)$~SNe are able to overcome the low probability of a coincident Fermi-LAT detection from the direction of an individual exploding SN.

\section{E.~Supplementary information on the statistical analysis}

\begin{figure*}
    \centering
    \includegraphics[width=1.\textwidth]{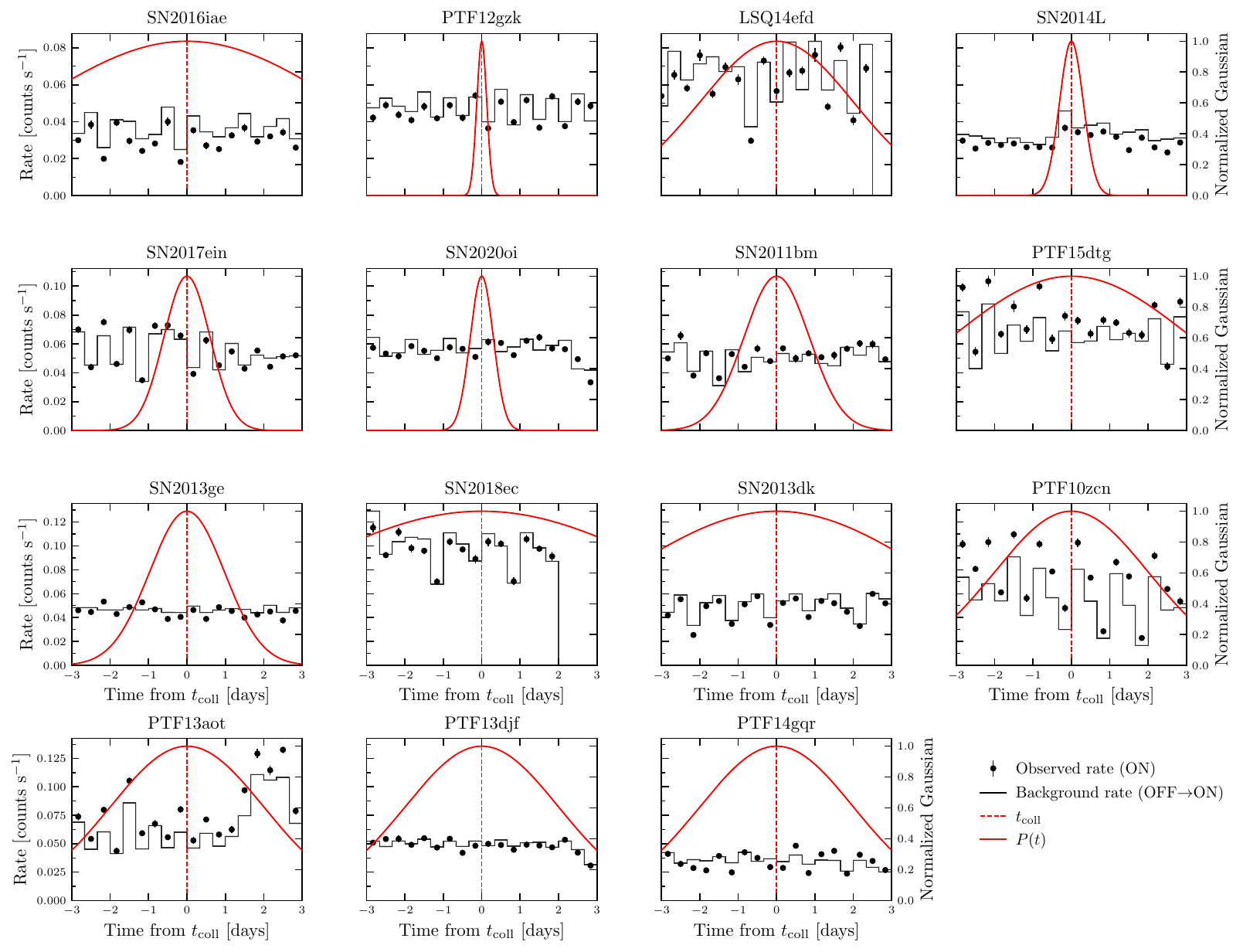}
    \caption{Light curves for each SN within the 6-day signal window, overlaid with the normalized probability distribution $P(t)$, modeled as a Gaussian centered at $t_{\rm coll}$ with a width corresponding to the explosion-time uncertainty. For clarity, the data are shown with an 8-hour binning. The full analysis uses a 20-day signal window for SN2013dk, SN2018ec, and SN2016iae. } 
    \label{fig:lightcurves}
\end{figure*}

\begin{figure}
    \centering
    \includegraphics[width=\columnwidth]{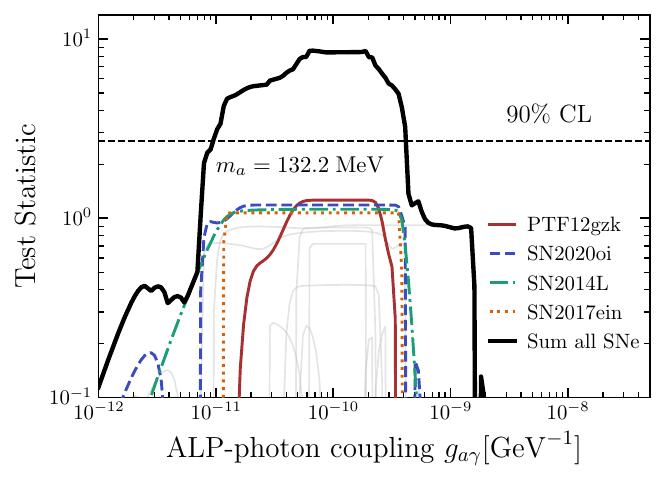}
\caption{Accumulated TS for all SNe in our sample as a function of the ALP–photon coupling $g_{a\gamma}$, with the most significant SNe highlighted in color and the less contributing ones shown in gray in the background. The ALP mass is fixed at $m_a = 132.2$~MeV. The total TS is displayed in black and compared to the threshold corresponding to the $90\%$ CL constraint.}\label{fig:TS_all_SNe}
\end{figure}

We now discuss in detail the statistical analysis we have performed. We consider a sample of 15 Type Ic SNe with explosion times determined by fitting their optical signal to estimate the moment of the collapse $t_{\rm coll}$ and its error \cite{Meyer:2020vzy, YoungSupernovaExperiment:2021fur, Barbarino:2017mes, Prentice:2018ual, Kankare:2021fzf, Tinyanont:2016vbz}. To estimate the background in a data-driven manner \cite{Muller:2023pip, Muller:2023vjm}, we define the background window $v_0$ as the interval from two years to three days prior to $T_0$, and the signal window $v_1$ as the interval spanning from three days before to three days after $T_0$. Although the ALP-induced signal is expected to be burst-like, uncertainties in the explosion times (ranging from a few hours to several days) necessitate such wide windows. Pass 8 photon events with energies between 30 MeV and 5 GeV are selected within $15^{\circ}$ of each SN position (Table~\ref{tab:SNe_sample}), subject to quality cuts (\texttt{DATA\_QUAL>==1\&\& LAT\_CONFIG==1}) and zenith angle $<100^{\circ}$ to reduce contamination from the Earth’s limb, detector artifacts, and cosmic-ray misclassification.

To estimate the background in a fully data-driven manner we proceed as follows \cite{Muller:2023pip, Muller:2023vjm}. The analysis was performed with the Fermitools (v2.2.0) and fermitools-data (v0.18), distributed via the Fermi Science Support Center~\cite{fermitools}. The LAT photon data are first binned in time with the \texttt{gtbin} routine using a fixed bin size $\Delta t = 1\, \mathrm{h}$, which yields the photon counts in each bin $N_i$. For each SN $i$ and temporal bin $\alpha$, we compute the exposure $\varepsilon_{i,\alpha}$ with \texttt{gtexposure} using the Fermi--LAT instrument response functions \texttt{P8R3\_TRANSIENT020\_V3}. We then slice this exposure into energy bins to obtain an energy-dependent quantity $\varepsilon_{i,\alpha}(E)$, which represents the time-integrated effective area as a function of energy in units of $\mathrm{cm}^2\,\mathrm{s}$. The effective observing time in the bin, \(T_i\), is taken from the Good Time Intervals (GTIs). Dividing the exposure by this live time yields the bin-averaged, energy-dependent effective area $A_{\mathrm{eff},i,\alpha}(E)$. We define the background window $v_0$ as the interval $(T_0 - 2\,\mathrm{yr},\,T_0 - 3\,\mathrm{d})$ and the signal window $v_1$ as $(T_0 - 3\,\mathrm{d},\,T_0 + 3\,\mathrm{d})$, where $T_0$ denotes the estimated explosion epoch. In the cases of SN2013dk, SN2018ec and SN2016iae we used $v_1$ as $(T_0 - 10\,\mathrm{d},\,T_0 + 10\,\mathrm{d})$ given the large uncertainty in the explosion time. Summing over the bins contained in each interval gives the total counts and exposures, namely $N_{\mathrm{off}, i} = \sum_{\alpha \in v_0} N_{i, \alpha}$ and $E_{\mathrm{off}, i} = \sum_{\alpha \in v_0} \varepsilon_{i, \alpha}$ for the background, and $N_{\mathrm{on}, i} = \sum_{\alpha \in v_1} N_{i, \alpha}$ and $E_{\mathrm{on}, i} = \sum_{\alpha \in v_1} \varepsilon_{i, \alpha}$ for the signal. The expected background contribution in the signal window $v_1$ is then obtained by exposure–weighted scaling, $B_{i, \alpha} = (N_{\mathrm{off}, i}/E_{\mathrm{off}, i})\,E_{\mathrm{on}, i, \alpha}$, which ensures that differences in live time and pointing conditions between $v_0$ and $v_1$ are properly accounted for.

For each SN $i$, both $v_0$ and $v_1$ are divided into 1-hour bins, indexed by $\alpha$. Tests with shorter binning showed no significant improvement in sensitivity. The collapse time $t_{\rm coll}$, at which the gamma-ray flash begins, is modeled as a Gaussian distribution $P(t_{\rm coll}),dt_{\rm coll}$ centered on the optically inferred explosion epoch, with width given by its uncertainty (See Figure~\ref{fig:lightcurves} and Table~\ref{tab:SNe_sample} of the SupM). We neglect here the delay between collapse and visible explosion due to the shock wave traversing the stellar mantle, typically of less than a minute, quite smaller than the uncertainty on the explosion time itself.

We assume the gamma-ray flash from ALP decay to happen within a single bin. Then for each bin $\alpha$ relating to the SN $i$, we can define the probability of the gamma-ray flash as $P_{i,\alpha}=\int_{\alpha} dt_{\rm coll} P(t_{\rm coll})$, where the integral is done only within the duration of the $\alpha$-th bin. Here $t_{\rm coll}$ marks the collapse time of the star, roughly simultaneous with the gamma-ray flash. For many of these bins, the experiment is not actually sensitive to the region of the sky where the SN is located. Let us denote by $\overline{\alpha}$ the indices of the bins in which the experiment was sensitive. Then the probability that the gamma-ray flash happens within a time frame in which the experiment was observing, for SN $i$, is $f_i=\sum_{\overline{\alpha}} P_{i,\overline{\alpha}}$. In these bins, the gamma-ray flash would have produced a number of signal events $S_{i,\overline{\alpha}}(g_{a\gamma},m_a)$ which can be obtained by integrating the gamma-ray fluence over the effective area for each bin $A_{\mathrm{eff},i,\overline{\alpha}}$, inferred by dividing the exposure of each bin by the effective observed time
\begin{equation}
    S_{i,\overline{\alpha}}(g_{a\gamma},m_a)=\int dE_\gamma \frac{dN_\gamma}{dE_\gamma}(g_{a\gamma},m_a)\frac{1}{4\pi r_i^2}A_{\mathrm{eff}, i,\overline{\alpha}}(E_\gamma).
\end{equation}
Here $r_i$ is the distance of the $i$-th SN from Earth, which values are listed in Table~\ref{tab:SNe_sample}. 

We show explicitly for our entire suite of SNe the observed and background number of events, together with the assumed time distribution for the gamma-ray flash $P(t_{\rm coll})$ in Figure~\ref{fig:lightcurves}. We can now compare the observed number of events with the expected one, assuming the existence of ALPs causing a gamma-ray flash with the signal events $S_{i,\overline{\alpha}}$. The likelihood for each bin is assumed to be Poissonian, so that the overall likelihood for the $i$-th SN is
\begin{align}\label{eq:likelihood}
    \Pi_i=\Pi_i^0 \left[1-\sum_{\overline{\alpha}}P_{i,\overline{\alpha}}\left[1-e^{-S_{i,\overline{\alpha}}}\left(1+\frac{S_{i,\overline{\alpha}}}{B_{i,\overline{\alpha}}}\right)^{N_{i,\overline{\alpha}}}\right]\right],
\end{align}
where 
\begin{equation}
    \Pi_i^0=\prod_\alpha \frac{B_{i,\alpha}^{N_{i,\alpha}}e^{-B_{i,\alpha}}}{N_{i,\alpha}!}
\end{equation}
is the likelihood in the pure-background case and the term in parenthesis in Eq.~\eqref{eq:likelihood} is the likelihood ratio between the case with signal and the pure-background case. We can then introduce the log-likelihood-ratio with the pure-background case $\Lambda_i=-2\log(\Pi_i/\Pi_i^0)$, with the factor $-2$ introduced for convenience to match with the definition of test statistic (TS). The total log-likelihood ratio for all the SNe is additive $\Lambda(g_{a\gamma},m_a)=\sum_i \Lambda_i$, where we now mark explicitly the dependence on the coupling and mass of the ALP. We can now finally define our test statistic (TS) which is used to set constraints on the ALP-photon coupling; since we do not want to set upper bounds, but rather generic bounds on the ALP-photon coupling---in the trapping regime, the larger couplings are allowed, rather than excluded---we choose as a TS the log-likelihood-ratio with the best-fit case
\begin{equation}
    \chi(g_{a\gamma},m_a)=
    \Lambda(g_{a\gamma},m_a)-\Lambda(\hat{g}_{a\gamma}(m_a),m_a),
\end{equation}
where $\hat{g}_{a\gamma}(m_a)$ is the ALP-photon coupling that maximizes the likelihood for each value of $m_a$. According to Wilks' theorem~\cite{Wilks:1938dza}, in the asymptotic limit of many events, this quantity is distributed according to a chi-squared distribution with one degree of freedom, so that we may finally obtain the value of $g_{a\gamma}$ corresponding to a $90\%$ confidence level (CL) by requiring the threshold $\chi(g_{a\gamma},m_a)=2.71$.

This analysis is necessary to compare the signal assumption---a prominent gamma-ray flash which may happen with probability $P_{i,\overline{\alpha}}$ in a visible bin, or with a probability $1-f_i$ in an invisible bin---with the observation. However, in this presentation it may be so convoluted as to obscure its physical meaning. 
Let us now consider a simple, but representative, case to build intuition into the outcome of this analysis. Assume for simplicity that $S_{i,\overline{\alpha}}\gg B_{i,\overline{\alpha}}, N_{i,\overline{\alpha}}$, so that the signal, if happening within a time frame in which the experiment was sensitive, would certainly be seen. Then we simply have $\Pi_i=\Pi_i^0(1-f_i)$, and therefore $\chi=\Lambda=-2\sum_i \log(1-f_i)$---the best-fit hypothesis is of course the pure-background one. So even if the fraction of visible time $f_i$ is small, the TS can still grow large if multiple SNe are included in the set. Assuming $f_i\sim 0.3$, we find that already $\sim 6$ SNe would be able to exclude this possibility with our statistical procedure. In Table~\ref{tab:SNe_sample}, we show the concrete values of $f_i$ for each SN, validating our intuitive calculation here, as well as the TS for the individual SNe, which does not reach up to the 90\% CL threshold, together with the total one, which instead does provide an exclusion.

Since gamma rays from ALP decay at large couplings would produce a huge signal count in Fermi-LAT---provided that they escape from the progenitor star---the main challenge in our analysis is not to produce enough events in the detector, but rather to ensure that it was sensitive to at least some of the SNe in our sample at the time of the explosion. Generally, the explosion time is not known with precision; even the visible explosion time can be reconstructed only up to a relatively large uncertainty (for all of the SNe we consider this is collected in Table~\ref{tab:SNe_sample}). There is also in principle an additional uncertainty of less than a minute, due to the time it takes for the shock wave to propagate from the collapsed core to the progenitor surface; we generally neglect this delay as it is much narrower than the uncertainty on the visible explosion time.

The overall time distribution of the visible gamma-ray flash is shown in Fig.~\ref{fig:lightcurves} for all of the SNe considered here, together with the estimated flux level in the detector, and the average level of the background. This shows clearly that the collapse time can be uncertain by several days, and also gives an immediate impression of the flux level that must be overshoot for a detection; the typical background flux for all SNe is of the order of $\phi_{\mathrm{LAT}\, \gamma}\sim 10^{-5}\, \mathrm{cm}^{-2}\, \mathrm{s}^{-1}$. For comparison, an ALP with a coupling $g_{a\gamma}\sim 10^{-10}\,\mathrm{GeV}^{-1}$ produces a number of photons from ALP decays of the order of $N_\gamma\sim 10^{53}$ (see Fig.~\ref{fig:spectrum}); at a distance of roughly $10$~Mpc, and within a bin of time duration of an hour as we adopt here, the corresponding signal photon flux in the detector would be of the order of $\phi_\gamma\sim 10^{-4}\, \mathrm{cm}^{-2}\,\mathrm{s}^{-1}$. Our constraints reach couplings much larger than $10^{-10}\, \mathrm{GeV}^{-1}$, making it clear that the signal flux from ALP decays, if arriving in a moment in which the detector is active, reaches orders of magnitude above the background.

Indeed, the test statistic (TS) as defined in the main text for each of the SNe, shown in Fig.~\ref{fig:TS_all_SNe}, nicely confirms all of our qualitative expectations. For each individual SN, the TS vanishes at low couplings---the signal flux here is much lower than the background---and at high couplings---photons are trapped within the progenitor. In between, however, it quickly saturates to a constant value; in this range, the signal flux is much larger than the background, and the exclusion probability depends purely on $f_i$, the probability that the event happens while Fermi-LAT is sensitive. As discussed in the main text, the saturated TS for each SN is roughly equal to $\chi_i=-2\log(1-f_i)$; the values of $f_i$ in Table~\ref{tab:SNe_sample} nicely match with the maximum TS for each SN in Fig.~\ref{fig:TS_all_SNe}. For some of the SNe, the TS exhibits a somewhat more complicated behavior, with multiple peaks, presumably due to the presence of solutions that can accommodate slight excesses in the data sample. Overall, each individual SN saturates to a value of TS that is unable to provide an exclusion. However, the total TS, coming from the combined power of all SNe together, exceeds the 90\% CL threshold for exclusion, leading to the constraints shown in the main text.


\end{document}